\documentclass[11pt,a4paper]{article}
\usepackage[pdftex]{graphicx}
\usepackage[cmex10]{amsmath}
\usepackage{amsfonts}
\usepackage{amssymb}
\usepackage{amsthm}
\usepackage{siunitx}

\usepackage{algorithm}
\usepackage{algpseudocode}
\usepackage{color}
\usepackage[hidelinks=true]{hyperref}

\usepackage{enumerate}

\usepackage{subfigure}
\usepackage[font=footnotesize ]{caption}

\usepackage{multirow}
\usepackage{rotating}
\usepackage{makecell}
\usepackage{booktabs}
\usepackage{makecell}

\usepackage{anysize}
\marginsize{2cm}{2cm}{1.5cm}{1.5cm}

\usepackage{setspace}
\usepackage{parskip}
\usepackage{lineno}

\usepackage{authblk}
\title{Machine learning approaches to understand the influence of urban environments on human's physiological response}
\author[1]{Varun~Kumar~Ojha\thanks{Corresponding Author: Varun~Kumar~Ojha,~email: vkojha@ieee.org\\
Information Sciences 474, 154-169}}
\author[1]{Danielle Griego} 
\author[2]{Saskia Kuliga}
\author[2]{Martin Bielik} 
\author[1]{Peter Bu\v{s}} 
\author[1]{Charlotte Schaeben} 
\author[1]{Lukas Treyer}
\author[1]{Matthias Standfest}
\author[2]{Sven Schneider}
\author[3]{Reinhard K\"{o}nig} 
\author[2]{Dirk Donath}
\author[1]{Gerhard Schmitt}

\affil[1]{Chair of Information Architecture, ETH Z\"{u}rich, Switzerland}
\affil[2]{Chair of Computer Science in Architecture, Bauhaus University, Weimar, Germany}
\affil[3]{Computational Architecture, Bauhaus University, Weimar, Germany}

\date{}
\begin{document}
	\onehalfspacing
	\maketitle
	\begin{abstract}
	This research proposes a framework for signal processing and information fusion of spatial-temporal multi-sensor data pertaining to understanding patterns of {\color{black}humans} physiological changes in an urban environment. The {\color{black}framework includes} signal frequency unification, signal pairing, signal filtering, signal quantification, and data labeling. Furthermore, this paper contributes to {\color{black}human-environment interaction} research, where a field study to understand the influence of environmental features such as varying sound level, illuminance, field-of-view, or environmental conditions on humans' perception was proposed. In the study, participants of various demographic backgrounds walked through an urban environment in Z\"{u}rich, Switzerland while wearing physiological and environmental sensors. Apart from signal processing, four machine learning techniques{\color{black},} classification, fuzzy rule-based inference, feature selection, and clustering{\color{black},} were applied to discover relevant patterns and relationship between the participants' physiological responses and environmental conditions. The predictive models with high accuracies {\color{black}indicate} that the change in the field-of-view {\color{black}corresponds} to increased participant arousal. Among all features, the participants' physiological responses were primarily affected by the change in environmental conditions and field-of-view. 
		
	~
	\end{abstract}
\begin{doublespace}
\section{Introduction}
\label{sec:intro}
Understanding {\color{black}influence of the} environmental conditions on human perception is complex. Various {\color{black}environmental features e.g., sound level, temperature, and illuminance affect our senses}. Therefore, we adopted enhanced measurement and analysis techniques to define and measure what influences citizens in dynamic urban environments. The environmental features measured in this research include sound level, dust, temperature, humidity, illuminance and the field-of-view since they influence a person's sense {\color{black}that, in this research, was represented by the physiological state of a person, which was measured through electro-dermal activity (EDA)}. With {\color{black}the} advent of technology, researchers {\color{black}explore} the utility of sensor-based physiological data in real-world scenarios. Thus, researchers now have a means to explore how environmental features can affect individuals' physiological response-based perceptual quality and overall experience~\cite{mavros2016}. How to capture and define such {\color{black}a perceptual quality} is an ongoing research topic in Cognitive Science and Behavioral Science~\cite{kuliga2017,varela2017}. 

This research presents a controlled study{\color{black},} conducted in Z\"{u}rich, Switzerland{\color{black},} to acquire {\color{black}data on humans physiological responses and environmental conditions}. In the study, 30 participants were asked to walk through an urban environment, while equipped with wearable sensor devices~\cite{griego2017}. The {\color{black}study} was designed to address the following research questions:
\begin{enumerate}[\quad(a)]
	\small
	\onehalfspacing
	\item Can we predict the physiological responses of participants based on particular environmental conditions? 
	\item Can we infer the relationship between the physiological responses and the environmental conditions?
	\item What are the most significant environmental features affecting the participants' physiological responses?  
	\item What are the patterns in the environmental conditions, for which the participants exhibit aroused and normal physiological responses?
\end{enumerate}

The {\color{black} features of the data} were {\color{black}recorded through} devices and sensors at varying {\color{black}frequencies}, which had both temporal and spatial properties. {\color{black}The features} had a temporal property due to continuous recording{\color{black}, and the features had spatial characteristics because the recording's association with the change in locations--global positioning system (GPS)}. Hence, in this research, we proposed a framework that perform signal preprocessing, signal filtering, signal quantifications, data fusion, and data labeling {\color{black}to answer the defined research questions}. 

Machine learning based techniques have been successfully applied for knowledge mining and pattern recognition in various real-world situations~\cite{rodriguez2016,witten2016} since they are useful in identifying the underlying patterns within data~\cite{alpaydin2014,ojha2017}. Thus, we formulated the processed {\color{black}data} such that four state-of-the-art machine learning techniques, classification, fuzzy rule-based inference, feature selection, and clustering, were applied for discovering patterns in the participants' physiological responses related to the urban environmental conditions. 

The first step in this research was to assess the predictability of participants' perception (physiological responses) of the urban environment. Thus, a ten-fold cross-validation was performed on a reduced error-pruning tree (REP-Tree) classification model~\cite{quinlan1987}. Following the classification approach, a fuzzy rule-based learning inferential model was built{\color{black},} using fuzzy unordered rule induction algorithm (FURIA)~\cite{huhn2009}{\color{black},} to investigate the relationship between the urban environmental features and the physiological response measures. Subsequently, the importance of various urban environmental features was analyzed by applying backward linear feature elimination filter (BFE)~\cite{maldonado2014}. Furthermore, self-organizing map (SOM)~\cite{kohonen1990} was applied to visualize the impact of urban environment features on participants' physiological responses. {\color{black}In the final step}, a method for referencing GPS location (geo-location) to compute mean physiological response across all participants {\color{black}was} developed. 
Since various methods were involved in data processing, additional graphics and multimedia can be found on the project website~\cite{esum2018}. 

{\color{black}In summary, following are three} essential contributions of this research:

\begin{enumerate}[\quad(a)]
	\onehalfspacing
	\item a field study design to understanding human perception of the  urban environment; 
	\item {\color{black}a framework design comprising} signal processing, signal quantification, and data fusion methods that {\color{black}invokes a novel of approach in physiological data quantification}; 
	\item a comprehensive analysis using four machine learning methods to discover the patterns which are crucial to our understanding of human perception in urban settings. 
\end{enumerate}
	 
We organized this paper into {\color{black}seven} Sections. Section~\ref{sec:hpue} {\color{black}places this research in the context of literature and describes the experimental procedure}. Section~\ref{sec:knowledge} describes signal preprocessing, multi-sensor information fusion, and machine learning techniques in detail. Section~\ref{sec:results} is devoted to explaining the obtained results followed by a comprehensive discussion in Section~\ref{sec:disc}. {\color{black}The challenges and opportunity of the research are presented in Section~\ref{sec:challanges}, and} Section~\ref{sec:con} concludes the findings of this research.

\section{Human perception of the urban environment}
\label{sec:hpue}

\subsection{Literature review}
The process of measuring physiological data as an indicator of human perception is complex, particularly in real-world application since perception can be influenced by various factors~\cite{bell2001}. However, physiological pattern recognition can derive significant evidence about human perception~\cite{picard2001}. Similar to our research, Picard et al.~\cite{picard2001} focused on physiological sensor data, specifically skin conductance, and they related high and low arousals as positive and negative biological reactions. Also, Picard et al.~\cite{picard2001} focused on the collection and filtering of the physiological data to construct good quality data void of failure and corrupt signals. They formulated physiological data so that a k-nearest-neighbor classifier can predict human's physiological arousal-based perception. Krause et al.~\cite{krause2003,krause2006}, on the other hand, used wearable device data, including physiology based sensor data (galvanic skin response), to identify user's state in terms of physiological and activity context using {\color{black}SOM} based clustering. Specifically, they performed unsupervised learning to classify sensor data to determine the context from which the signals were generated. 

In Wang et al.~\cite{wang2014}, pattern recognition and classification of physiological sensor signals were performed by first decomposing signals into its constituent features and by applying support vector machine to classify negative and positive emotion labels. Here, the label associated with the signals were predefined during the experiment by exposing the participant to  negative and positive environments during the recording of signals. Rani et al.~\cite{rani2006} performed an empirical study of four machine learning techniques: k-nearest neighbor, regression tree, Bayesian network and support vector machine for the recognition of the emotional state from physiological response data. They performed signal processing to evaluate features from the physiological data and labeled them with the emotional state reported by the participants. 

Since we investigate ``cause and effect'' between the environmental conditions and the human's perception, unlike Wang et al.~\cite{wang2014} and Rani et al.~\cite{rani2006}, we performed signal processing on the physiological data to evaluate skin conductance response (SCR) arousals~\cite{xia2015}. Subsequently, we  assigned labels to signal fragments based on the degree of arousal within a specified time. While doing this, we considered physiological data as the output in the classification model and the signals from the environment as the inputs. Whereas, Wang et al.~\cite{wang2014} and Rani et al.~\cite{rani2006} considered features of the processed data as the inputs and the reported environment as the output. Our approach{\color{black}, to first determine arousal level} was adopted {\color{black}because of} the complexities of the urban environment {\color{black}and because} we cannot accurately consider an urban environment to be positive or negative towards the perceptual quality of a participant. Thus, we labeled environmental conditions as the positive and negative by considering physiological data as the target in the classifier's training.  
     
Ragot et al.~\cite{ragot2017} found that the physiological response signals from the Empatica E4 wearable device were closely comparable to laboratory-based measurement devices. They also found that the data from such wearable devices could be used to train a support-vector-machine classifier to recognize the participants' emotional state.  Similarly, Poh et al.~\cite{poh2010} confirmed that EDA data from wearable devices is comparable to laboratory devices and the data are a valid physiological measure. {\color{black}Hence, was our approach in this study to employ Empatica E4 to perform physiological measure.}


\subsection{Study design and measurements}
\label{subsec:study}
\begin{figure}
	\centering
	\subfigure[]
	{
		\includegraphics[scale=0.35]{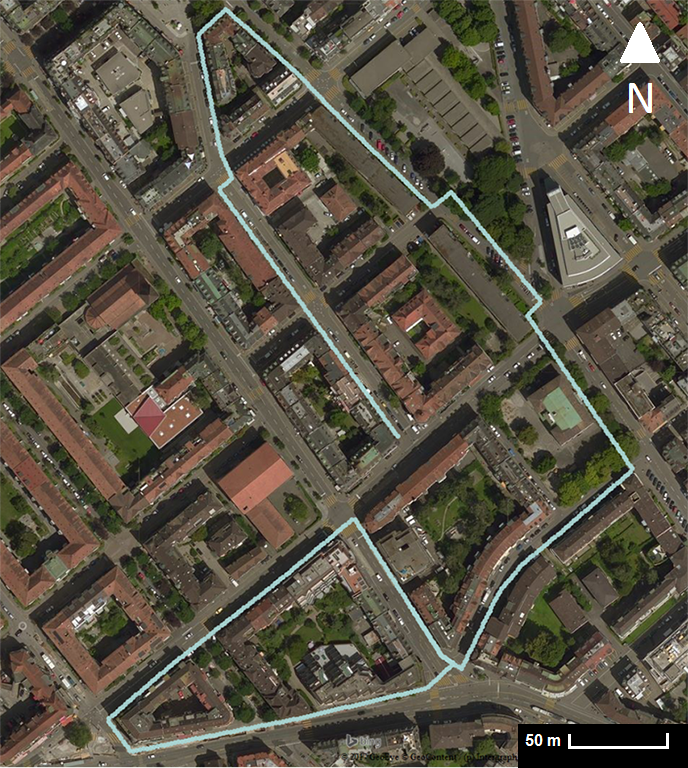}
		\label{fig_study_path}%
		
	} 
	\subfigure[]
	{
		\includegraphics[scale=0.35]{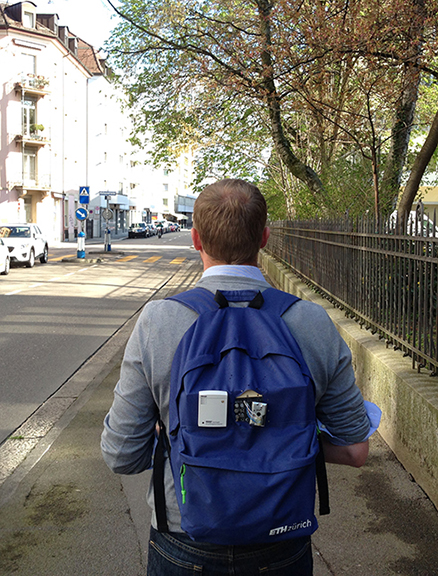}
		
		\label{fig_p_1}
	}
	\caption{(a) Study neighborhood marked with the participants' walking path (Wiedikon, Z\"{u}rich, Switzerland); (b) a participant with sensor backpack~\cite{esumbackpack2017}.}
\end{figure}

We designed a study to understand the general pattern(s) of human perception related to events which occur in a dynamic urban environment. An event indicates the change in the environmental condition, and also, a sample of the measured environmental data. 
As a case study, we selected a neighborhood in Z\"{u}rich, Switzerland (Fig.~\ref{fig_study_path}), and invited participants to take a leisure walk on a predetermined path (Fig.~\ref{fig_p_1}). {\color{black}The} participants were equipped with a {\color{black}``sensor backpack~\cite{esumbackpack2017}'' and an Empatica E4 wearable device~\cite{empatica2018}}. The \SI{1.3}{\kilo\meter} walking path was carefully selected {\color{black}, which covered} a diverse urban scenario~\cite{griego2017}, e.g., spacious and narrow streets, green and urban areas, and loud and quieter locations. 

Our sensor kit~\cite{esumbackpack2017} measured the changes in sound level (decibel, \si{dB}), the amount of dust (\si{mg/m^3}), temperature ($^{\circ}$C), relative humidity (\%), and illuminance (\si{\lux}). We also calculated field-of-view based on the GPS information and spatial configuration of the neighborhood. The field-of-view is formerly described as the Isovist descriptor, which refers to the open space a person can view from a single vantage point~\cite{benedikt1979}. Since participants were walking in a forward direction, we considered \ang{180} field-of-view with a distance of \SI{100}{\meter}. Subsequently, the Isovist descriptor for each participants' walk was measured by drawing a polygon around the participants' \ang{180} field-of-view at their specific GPS locations. From this, the following measures of the Isovist polygons were calculated: 
Area--{\color{black}polygon's surface area}; 
Perimeter--{\color{black}polygon's perimeter length}; 
Compactness--{\color{black}the} ratio of area to {\color{black}the} perimeter (relative to an ideal circle); {\color{black}and} 
Occlusivity--{\color{black}the} length of occluding edges.

The EDA measures the individuals' physiological state~\cite{braithwaite2013}, which was recorded using Empatica E4 wearable device, similar to studies by~\cite{empatica2018,esum2018,garbarino2014}. We placed the wearable device on participants' non-dominant hand and let it adjust for 10 minutes according to Empatica guidelines~\cite{empatica2018}. The data were recorded {\color{black}on} the Empatica website and corrected for motion artifact~\cite{empatica2018}. {\color{black}The EDA measure (physiological response) was a time-series signal and has temporal dependencies.} The sensor backpack{\color{black}, on the other hand,} was designed to capture the contextual-based events that occur in an urban environment. In the context of this study, an event is non-temporal since an event is dependent on the instance of {\color{black}its} observation. Therefore, the continuous signals recorded for environmental features and the continuous signals recorded for participants' physiological responses were quantified in two different manners (Section~\ref{subsec:data_qanti}). Moreover, since the recorded signals were associated with the geographical location, they also had spatial properties. The primary infrastructure of the urban environment and season (April 2016) were uniform. However, inherent diversity occurred from different experiment days, time-of-day, and participants demographic background. The {\color{black}data} for both environment measures and corresponding participants' physiological response measures {\color{black}are} summarized in Table~\ref{tab_study}.

\begin{table}
	\centering
	\renewcommand{\arraystretch}{1.2}
	{\footnotesize 	
	\caption{Measured features in the study.}
	\label{tab_study}
	\begin{tabular}{cccc}
		\toprule
		Data & Type & Features (sensors) & Frequency (Hz)\\
		\hline
		\multirow{9}{4cm}{\centering Urban environment (indicate the changes in the urban environmental condition during a participants' walk) } & \multirow{9}{2cm}{\centering Spatial (in the context of this study)} & GPS position (Latitude and Longitude) & 1.0 \\
		& & Sound level (\si{dB}) & 0.4 \\
		& & Dust in air (\si{mg/m^3}) & 0.4 \\
		& & Environmental temperature ($^{\circ}$C) & 1.0 \\
		& & Relative humidity (\%) & 1.0 \\
		& & Illuminance (\si{\lux}) & 1.0 \\
		& & \multirow{3}{6cm}{\centering Participants field-of-view (computed  based on GPS position): Area, Perimeter, Compactness, Occlusivity } & \multirow{3}{*}{-~-} \\
		& &  &  \\
		& &  &  \\
		\multirow{2}{5cm}{\centering Human perception (participants' physiological response)} & \multirow{2}{*}{Spatial-temporal} & \multirow{2}{*}{Electro-dermal activity (EDA)} & \multirow{2}{*}{4.0}  \\
		&  & 	 & \\
		\bottomrule
	\end{tabular}}
\end{table}	

\section{Methodologies}
\label{sec:knowledge}

\begin{figure}[!h]
	\centering
	\includegraphics[width=0.95\columnwidth]{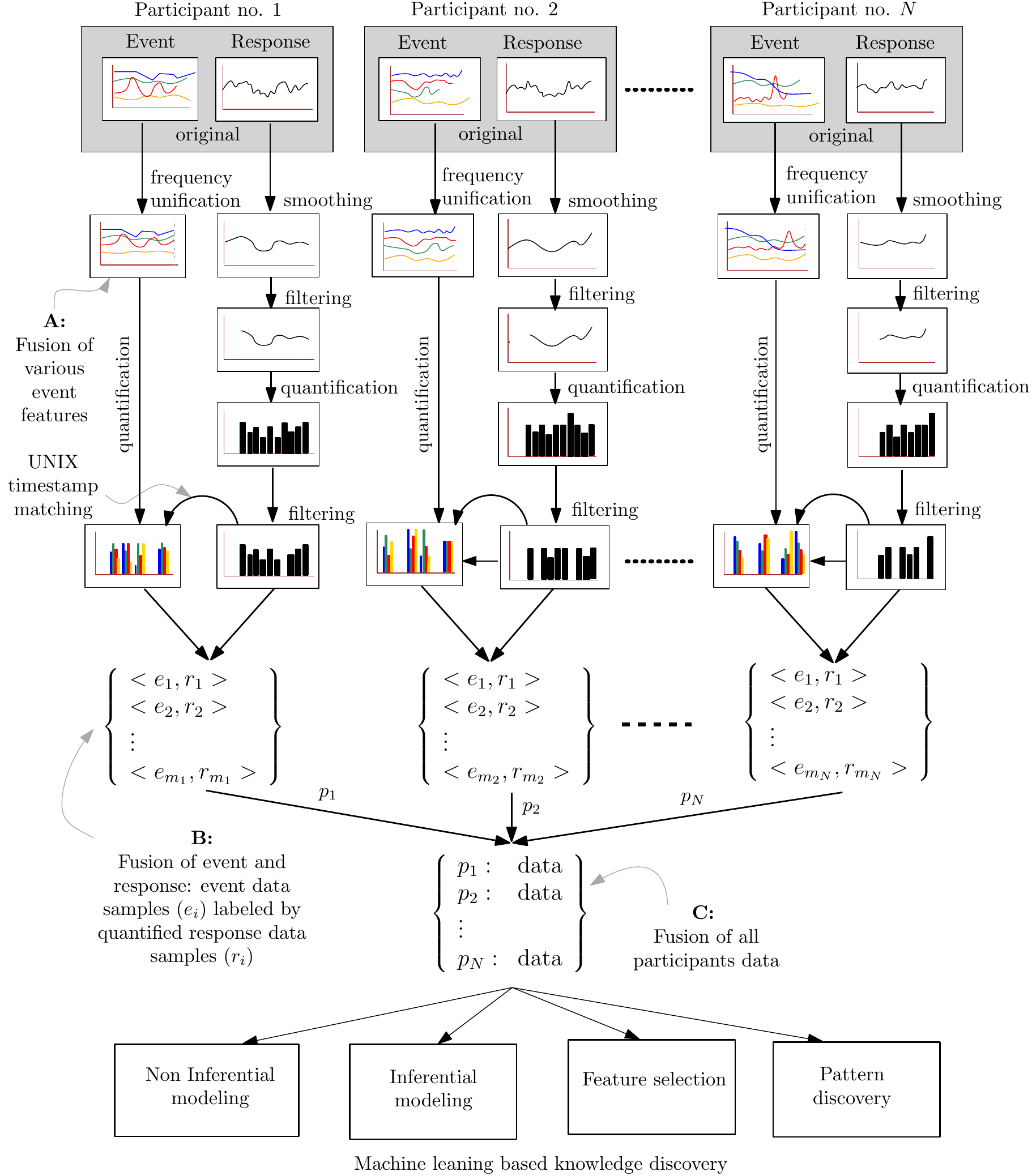}
	\caption{Information fusion and knowledge mining framework.}
	\label{fig_main_framework}
\end{figure}

A comprehensive signal processing and data-preprocessing framework {\color{black}was} proposed in order to apply select machine learning methods. Fig.~\ref{fig_main_framework} illustrates the framework and {\color{black}describes} how it was used for information fusion and knowledge mining approaches. Here, $e_i$  and  $r_i$   indicate $ i $-th quantified \textit{event} (a sample in the quantified environmental data) and \textit{response} (a sample in the quantified physiological response data) respectively. The variable $m_j$  for $j\in\{1,2,\ldots,N\}$ indicates the total number of samples belonging to the $j$-th participant $p_j$. The information{\color{black}, therefore, was} fused in three stages:  
\begin{enumerate}[\quad(a)]
\onehalfspacing
	\item Each participants' event-based data ($ e $) are collected from five sensors{\color{black}, which were re-sampled to a unique frequency and samples were aligned as per with on their time} (Fig.~\ref{fig_main_framework}, mark ``A'').

	\item The environment and response data from each participant were independently cleaned, filtered, and quantified. Each participants' quantified event and response data were fused (paired) by assigning a quantified response $r_i$ to event $e_i$ (Fig.~\ref{fig_main_framework}, mark ``B''). 

	\item The paired participants' data were then stacked (Fig.~\ref{fig_main_framework}, mark ``C''). 
\end{enumerate}

The three-stage information fusion approach produced the compiled dataset, which was fed to select machine learning techniques. For each machine learning technique, the compiled dataset (Fig.~\ref{fig_main_framework}, mark ``C'') was arranged and configured as per the techniques' requirements and objectives.   

\subsection{Signal processing}
\label{subsec:data_clean}
\subsubsection{Frequency unification}
\label{subsec:ferq_uni}
The environmental features sound and dust were collected at 0.4 Hz frequency; while GPS position, temperature, humidity, and illuminance were collected at 1 Hz frequency (Table~\ref{tab_study}). Therefore, an up-sampling mechanism with a linear interpolation was applied to sound and dust data~\cite{blu2004} to unify the frequencies of the gathered data. All features were then aligned to the same timestamp, which was crucial to ensure that all sensor values belong to an exact event during the study.

\subsubsection{Signal filtering and smoothing}
\label{subsec:signal_filter}
The physiological response data (EDA signals) were kept at their original 4Hz frequency to maintain the information required for arousal detection from the physiological data. With close inspection, we found that some participants EDA signals were unusable and were discarded. The remaining (accepted) EDA signals were first smoothed and then filtered to remove artifacts as recommended in EDA  literature~\cite{braithwaite2013,chen2015}. 

\begin{figure}
	\centering
	\subfigure[]
	{  
		\includegraphics[width=0.43\columnwidth]{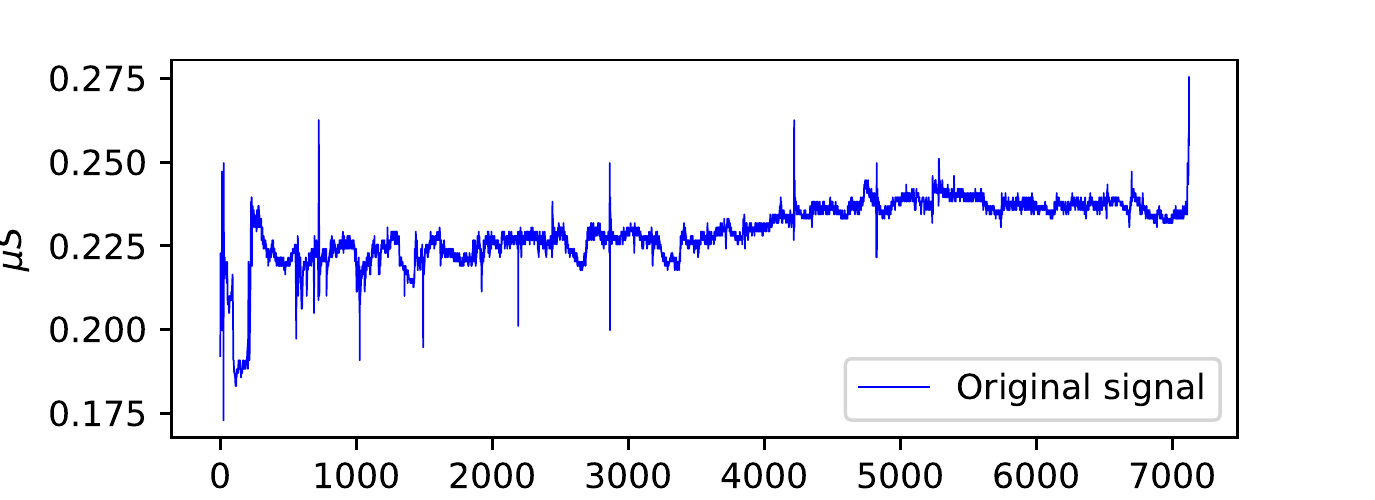}
		\label{fig_corr_1}%
	}
	\subfigure[]
	{    
		\includegraphics[width=0.4\columnwidth]{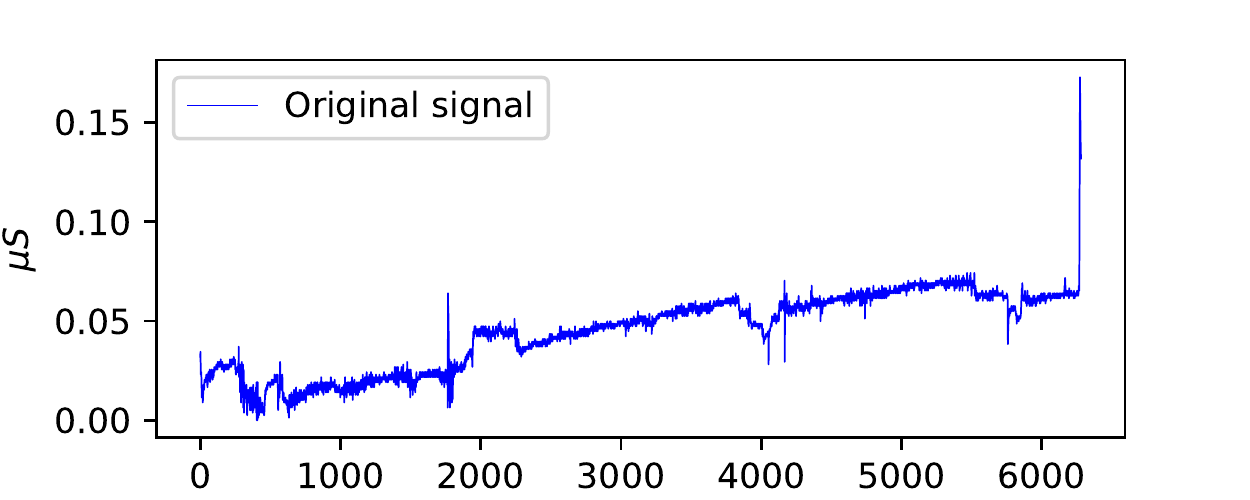}
		\label{fig_corr_2}%
	}
	
	\subfigure[]
	{   
		\includegraphics[width=0.4\columnwidth]{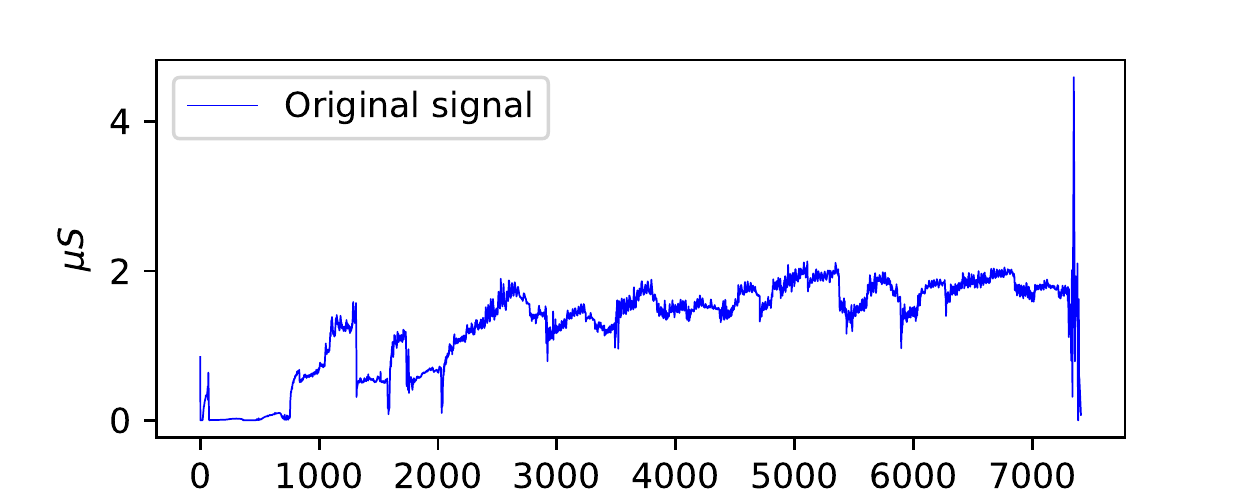}
		\label{fig_corr_3}%
	}
	\subfigure[]
	{   
		\includegraphics[width=0.425\columnwidth]{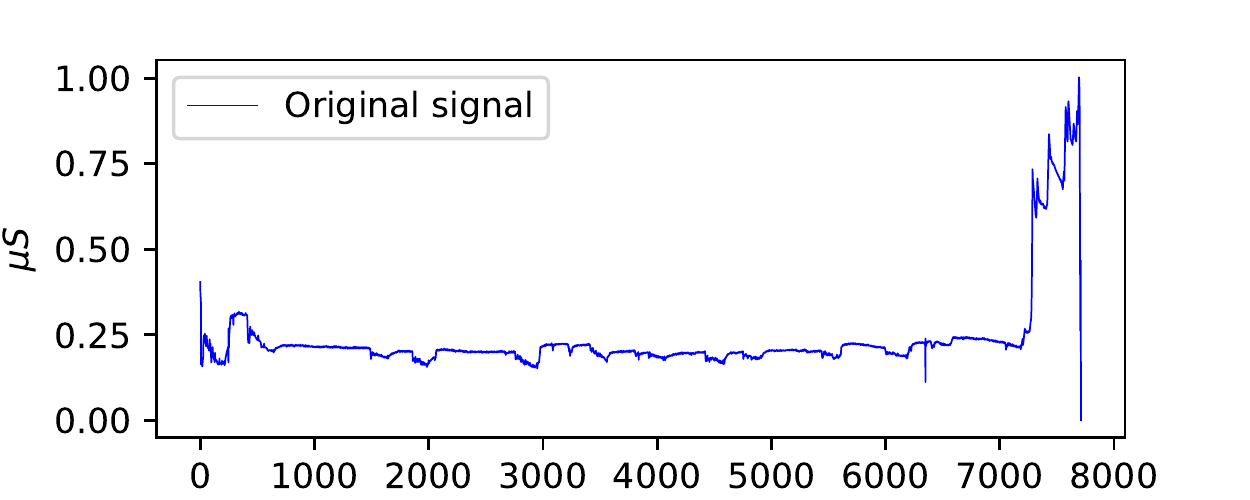}
		\label{fig_corr_4}%
	}

	\subfigure[]
	{   
		\includegraphics[width=0.45\columnwidth]{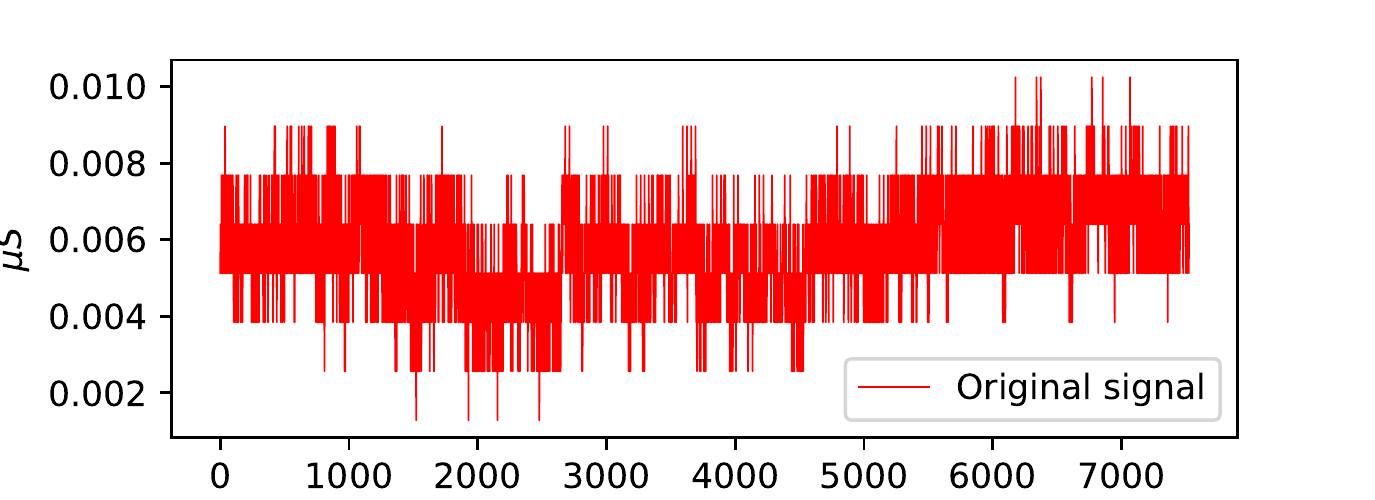}
		\label{fig_err_1}%
	}
	\subfigure[]
	{   
		\includegraphics[width=0.42\columnwidth]{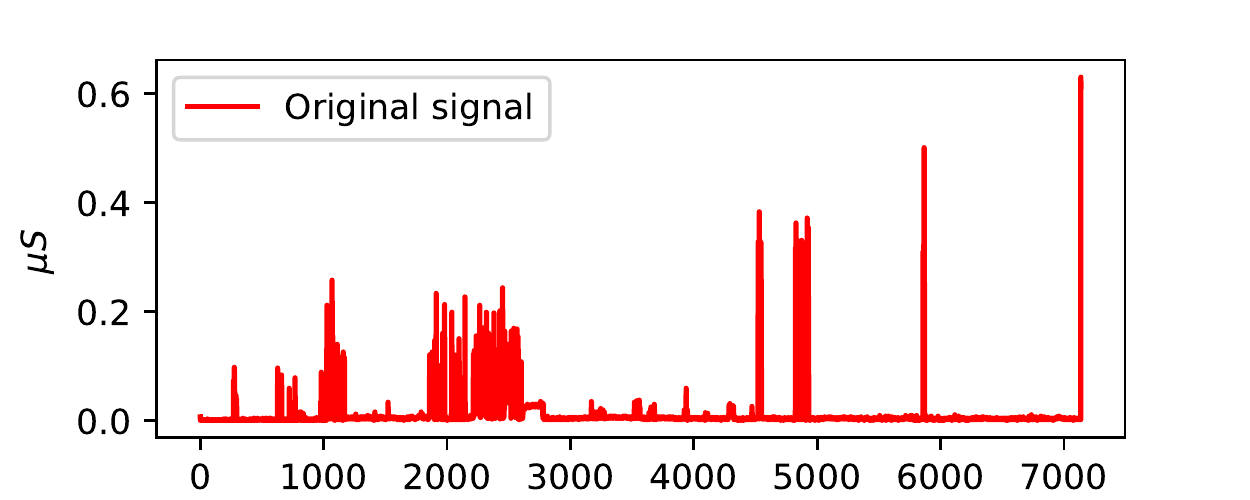}
		\label{fig_err_2}%
	}
	\caption{Signals in (a), (b), (c), and (d) are the most commonly found EDA signal profiles and considered for the analysis. Most commonly found error in signals are shown in (e) and (f).}
\end{figure}

\paragraph{Physiological data selection}
The EDA signals from 30  participants were analyzed by comparing the various ``profiles.'' The EDA signals from four types of uncorrupt EDA profiles shown in Figs.~\ref{fig_corr_1},~\ref{fig_corr_2},~\ref{fig_corr_3}, and~\ref{fig_corr_4} were considered for the data analysis. The EDA signals belonging to the two erroneous EDA profile types illustrated in Figs.~\ref{fig_err_1} and~\ref{fig_err_2} were discarded. In total 10 EDA signals were discarded. The erroneous EDA signal types were classified as: 
\begin{enumerate}[\quad(a)]
	\onehalfspacing
	\item Type-1 error, when EDA signal values only fluctuate between two values, i.e., the EDA signal behaved like a step function, and the signal may also contain a significant amount of sensor loss (no sensor response record).
	
	\item Type-2 error, when the majority of the sample values were zero (significant {\color{black}sensor response} loss), despite the otherwise normal fluctuations (correct sensor response) in EDA signal.
\end{enumerate}

\paragraph{Stationary Wavelet Transformation based smoothing}

\begin{figure}
	\centering
	\subfigure[]
	{
		\includegraphics[width=0.465\columnwidth]{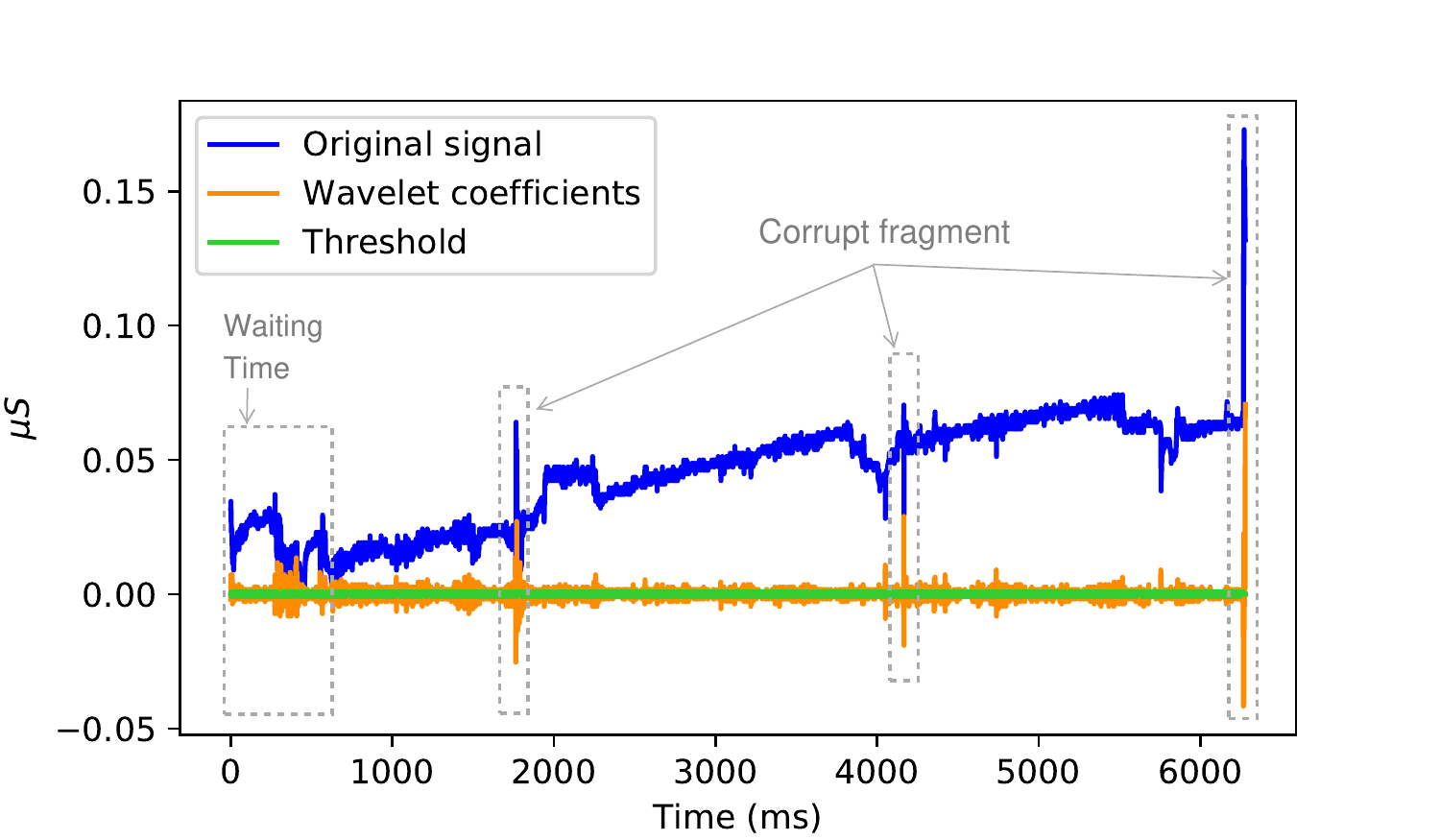}
		\label{fig_swt}%
		
	}
	\subfigure[]
	{
		\includegraphics[width=0.44\columnwidth]{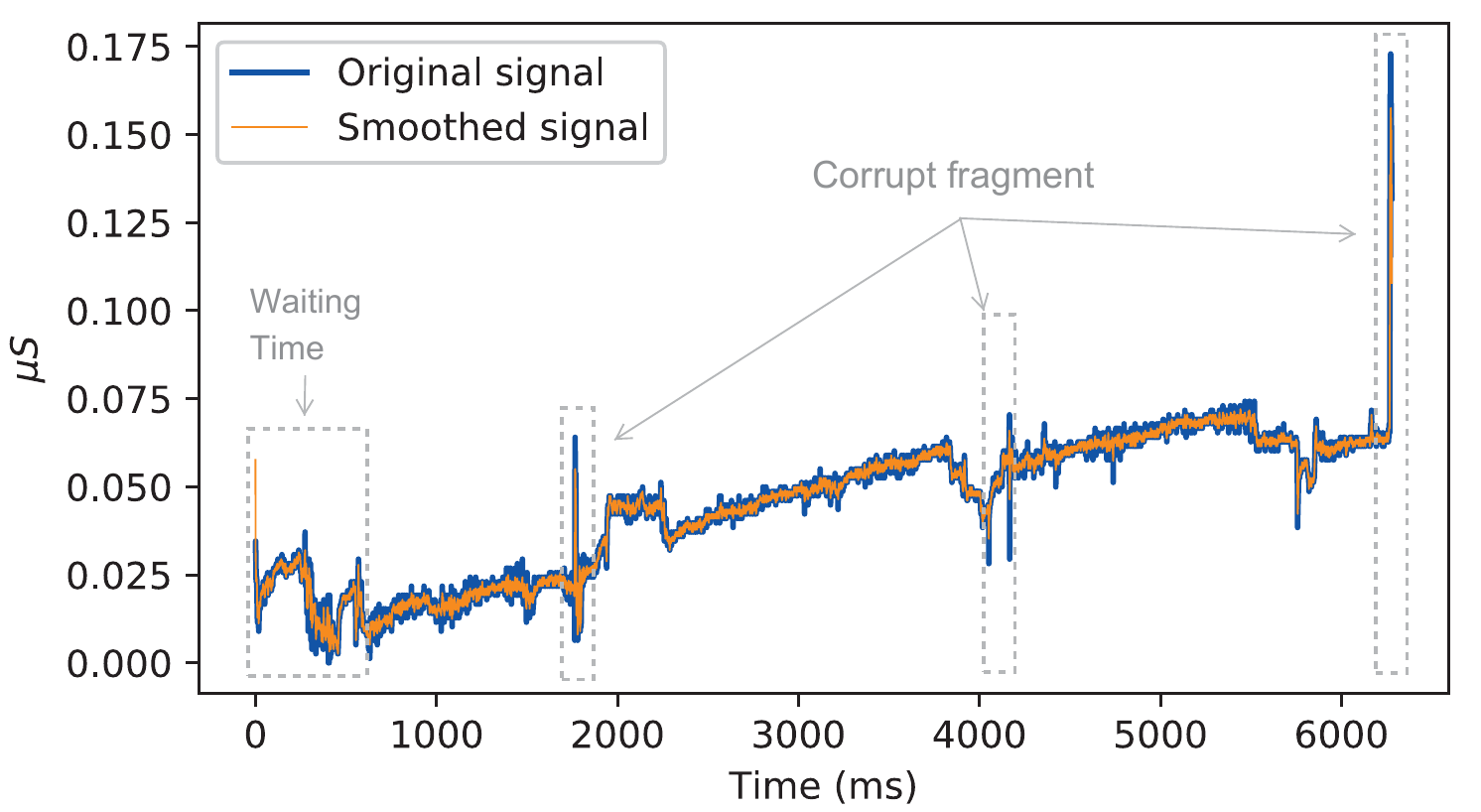}
		\label{fig_swt_trunk}%
		
	}
	\caption{Stationary Wavelet Transform based smoothing. (a) Wavelet transform of an original EDA signal using “Haar” wavelet, and smoothing by applying a threshold over wavelet coefficient. (b) Original and smoothed EDA signal with filtering of corrupt and unnecessary fragments.}
\end{figure}

After selecting EDA signals, they were smoothed by undergoing a Stationary Wavelet Transformation (SWT) and reverse SWT. Authors in~\cite{chen2015} suggested an adaptive method for SWT-based smoothing for EDA signals recorded for long periods (30 hours). In our study, EDA signals were recorded for 25--29 minutes. Therefore, we applied a one-level SWT and {\color{black}reverse-SWT} for smoothing. Each EDA signal was transformed using ``Haar'' as a mother wavelet in the SWT~\cite{molavi2012}. A one-level SWT transformation was performed on each signal; and on the obtained wavelet coefficients, a threshold of value $ \pm0.001 $ was applied to eliminate larger fluctuation in the signal. That is, the values of wavelet coefficients above $ +0.001 $ and below $ -0.001 $ were cut off (Fig.~\ref{fig_swt}). Finally, a reverse SWT was applied to the transformed signal to produce a smoothed signal (Fig.~\ref{fig_swt_trunk}).

\paragraph{Truncation of the unwanted signal fragments}
SWT based treatment to the EDA signals eliminated the large fluctuations from the signal. However, some sharp drops in signal (corrupt fragment) caused by {\color{black}artifact} were not filtered out completely. Thus, the corrupt fragments and participants' waiting time fragments of EDA signal were truncated from both original (raw) and smooth EDA signals. Fig.~\ref{fig_swt_trunk} is an example of such truncation. This process produced two EDA signals: original (original signal with filtering only) and smooth (original signal with both smoothing and filtering).

\subsection{Signal quantification and labeling}
\label{subsec:data_qanti}
Signal  quantification involved three steps: time-window marking, arousal detection, and data labeling. In fact, these are the critical steps in the fusion of the environmental data and the physiological response data. As shown in Fig.~\ref{fig_main_framework}, at first, physiological data were quantified, and then, the timestamp information was passed to the environmental data for its quantification. 

\subsubsection{Time window marking}
\begin{figure}
	\centering
	\subfigure[]
	{
		\includegraphics[width=0.9\columnwidth]{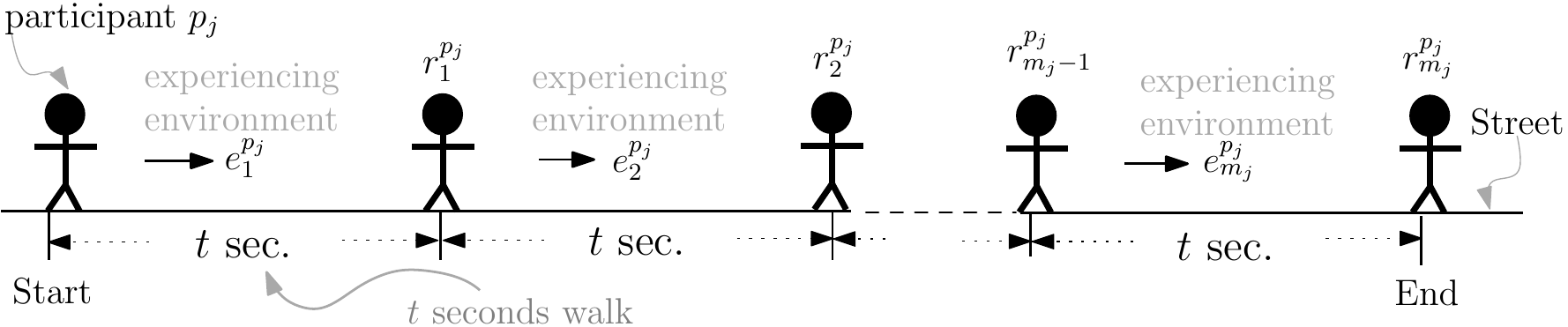}
		\label{fig_walk}%
	}
	
	\subfigure[]
	{
		\includegraphics[width=0.9\columnwidth]{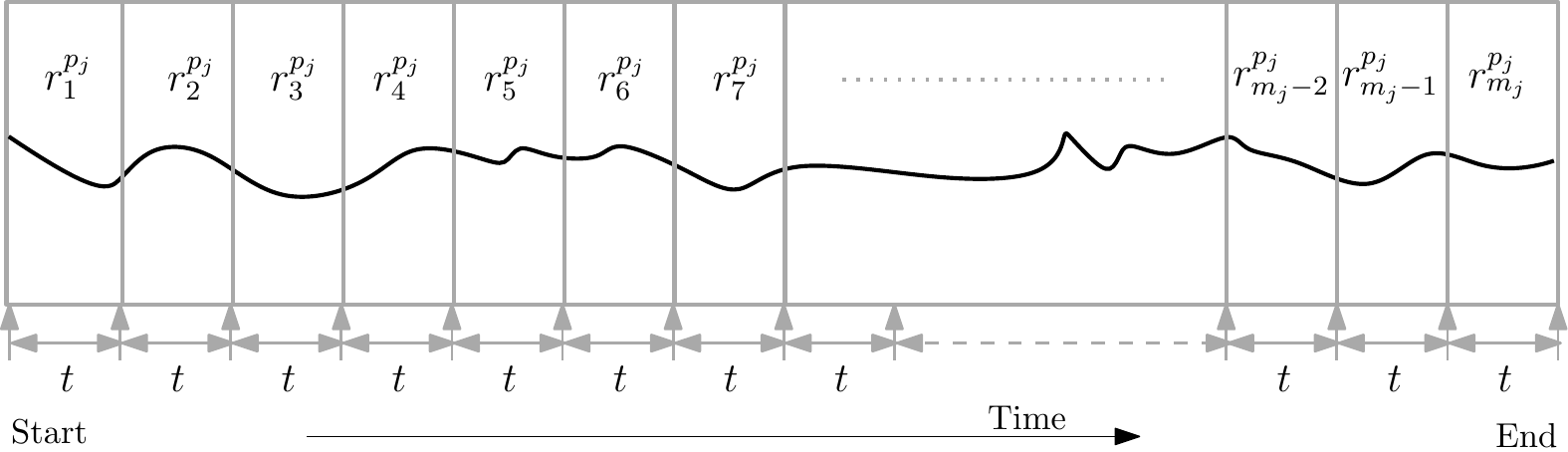}
		\label{fig_walk_response}%
	}
	\caption{(a) Timestamp is indicating Start and End of a participants' walk during the study. It illustrates the approach to quantify a {\color{black}participant's} physiological response and environmental experience data (b) Timestamp and time-window marking for an EDA signals (physiological response) at every $ t $ seconds for the detection of arousal $ r_i^{p_j} $ for $ i = 1 $ to $ m_j $. }
	\label{fig_time_mark}
\end{figure}

Each EDA {\color{black}signal's} timestamp information was compared with the timestamps recorded at various stages during a participants' walk.  Based on signal filtering shown in Fig.~\ref{fig_swt_trunk} and available timestamp information, the signal fragment  belonged to the walking duration---indicated by Start and End in Fig.~\ref{fig_walk}---were marked with a regular interval of time-window size $ t $ seconds. Such a time-window marking was crucial to our data analysis to observe participants physiological states in relation to their experience of the events occurring at a regular interval of $ t $ seconds (Fig.~\ref{fig_walk}). 

{\color{black}Therefore}, for each time-window, event $ e_i^{p_j} $ for $ i = 1 $ to $m_j$ experienced by participant $ p_j $ is a vector of the environmental features and was computed by averaging the values of signal fragment (environmental measurement) at the $ i $-th corresponding time-window. On the other hand, the participants physiological response $ r_i^{p_j} $  for $i = 1$ to $m_j$ upon experiencing event $ e_i^{p_j} $  was computed by an arousal detection method described in Section~\ref{sec:arousal_detection}. Additionally, the participants' field-of-view (Isovist descriptors: area, perimeter, occlusivity, and compactness) were computed at the start of each time-window. Thus, participant  quantified data $p_j$ had {\color{black}an} identically independent vector of environmental conditions (event $ e_i^{p_j} $) and a corresponding physiological state (response $ r_i^{p_j} $) for each time-window.

\subsubsection{Arousal detection (EDA)}
\label{sec:arousal_detection}
The level of arousal $ r_i^{p_j} $ in an EDA signal depends on identifying a specific signature (pattern) called skin conductance response (SCR) or arousal~\cite{benedek2010,braithwaite2013,choi2012,roth2012,taylor2015}. The state of arousal in an EDA signal is typically defined as a peak having a specific signature~\cite{braithwaite2013}. We processed the EDA signals using a skin conductance processing tool Ledalab~\cite{benedek2010}. Ledalab offers a continuous decomposition analysis (CDA) method for analyzing an EDA signal. In CDA, an EDA signal is decomposed into tonic skin conductance level (SCL) and phasic drivers SCR. 

We performed CDA on each EDA signal data---of each participant---by using the recommended settings in Ledalab~\cite{benedek2010}. {\color{black} That is, the signal's optimization procedure was performed two times, which automatically determined the optimization parameters for evaluating the number of significant SCR (nSCR) above a defined threshold 0.01$\mu$Siemens within a time-window.} We used nSCR, because we could not, in a theory-driven manner, define what stimulus (event) caused a change in participants ``physiological arousal state.'' Thus, we relied on a data-driven approach by analyzing phasic SCR, a non-specific fast changing EDA measure; i.e., the number of peaks in phasic skin conductance response measures nSCR to any kind of event for the given time-window. Therefore, the nSCR gave us the measures of $ r_i^{p_j} $ shown in Fig.~\ref{fig_walk_response}. 

\subsubsection{Data labeling}
\label{sec:datalabel}
When aggregating all participants data (Fig.~\ref{fig_main_framework}, mark ``C''), we observe that nSCR value for a time-window vary from 0 to 12. An nSCR value 0 indicate that, in a time-window, a participant had a normal physiological condition. On the other hand, an nSCR value greater than 0 for a time-window indicates that a participant experienced a state of arousal at least once in that time-window. Thus, for the labeling of each time-window---of each participant data---a binary-class label indicating a binary state of phasic nSCR $ r_i^{p_j} $ can be used, where

\begin{enumerate}[\quad(a)]
	\onehalfspacing
	\item class 0 is ``normal'' physiological response (``N''), i.e., an nSCR value equal to 0; and 
	\item class 1 is ``aroused'' physiological response (``A''), i.e., an nSCR value greater than to 0. 
\end{enumerate}

A multi-class classification was also used, in which case, aroused physiological response{\color{black},} ``A'' has two categories: class ``LA'' indicating low arousal response, i.e., {\color{black}$0 < \text{nSCR} < 6$} and class ``HA'' indicating high arousal response, i.e., {\color{black}$ \text{nSCR} \ge 6$}.  
A total of 6,057 samples and 9 input features were available in the compiled dataset for a time-window size $ t $ (quantification rate) of 5-seconds. In the compiled data, 3,491 samples belonged to the category ``N'' and 2,566 samples belonged to the category ``A,'' i.e., approximately {\color{black}60\% and 40\% of the samples respectively belong to ``N'' and ``A.''} Furthermore, in the multiclass classification, 2,079 samples were labeled ``LA'' and 487 samples were labeled ``HA.''

	
\subsection{Machine learning methods}
\label{subsec:know_mine}

\subsubsection{Non-inferential modeling}
\label{subsec:non_inf}
We build a predictive model consisting of the environmental features as the inputs, and binary ({\color{black}and} multiclass) quantified arousal level as the output using REP-Tree, which is a decision tree learner~\cite{quinlan1987}. In a decision tree, a tree-like predictive model is built, where the leaves represent the target (e.g., the class labels: ``N'' or ``A'') and the branches represent an observation for a feature (e.g., sound level) at a node. REP-Tree is a method applied to reduce the size of a decision tree, where it keeps pruning subtrees by replacing it with a leaf (a class label) as long as the error does not increase (i.e., the accuracy of the model does not decrease). 

We chose REP-Tree to build a predictive model because the algorithm constructs a decision tree, where each node makes a decision for a feature, and its specific value produces a particular class label. While making a predictive model, REP-Tree chooses the most significant features based on their contribution to the model's accuracy, which is advantageous for this problem since it is uncertain which environmental features influence physiological responses. For the validation of the model's predictive performance, we chose ten-fold cross-validation (10-fold CV). Section~\ref{sec:results} describes the test accuracies of 10-fold CV based REP-Tree training. 

\subsubsection{Inferential modeling}
\label{subsec:inf}
Contrary to non-inferential modeling, inferential modeling explains the relationships between the input features and the output feature.  A fuzzy rule-based inference system is capable of describing how independent environmental features are related to the dependent physiological response (phasic nSCR) feature. For this, we applied FURIA, which is a fuzzy rule-based classifier~\cite{huhn2009}.

Unlike conventional rule-based classifiers, FURIA gives a fuzzy rule~\cite{huhn2009}. FURIA produces fuzzy rules with operators $ \le $, $ = $,  {\color{black}and} $ \ge $; the operators define clear conditions for a feature's association with a class label (e.g., ``N'' or ``A''). FURIA also provides a range (e.g., $ x \rightarrow y $) indicating fuzziness in feature's condition, which may be considered as a soft boundary while associating a feature with a class label~\cite{huhn2009}. This ability was particularly useful in this study since we wanted to observe the specific values range of the environmental features that corresponded to a participants' state of arousal. For instance, we needed to determine for which particular sound level range, a participant experienced a state of arousal. Since FURIA fulfills this requirement, it was selected as the technique for inferential analysis. Interpretation of the obtained rules is described in Section~\ref{sec:results}.


\subsubsection{Feature selection}
\label{subsec:fs}
Feature selection is a process to determine the ability of each input feature {\color{black}to} predict the output. Moreover, feature selection involves making a model using a subset of features and testing its predictive accuracy. We applied backward feature elimination (BFE) method in this research for its ability to examine all possible combinations of feature subsets~\cite{maldonado2014}. BFE starts with all features in a set (in this case, it begins with 9 features) to build and test the model. Subsequently, BFE iteratively eliminates features one-by-one while 
propagating high accuracy feature subsets to the next iteration. Finally, 
BEF gives a list of subsets with their corresponding accuracies, from which a subset can be selected depending on the accuracy or the number of features required. 
In addition to REP-Tree, MLP~\cite{hagan1994} and SVM~\cite{chang2011} were used for a more comprehensive analysis in BFE. Therefore, the feature selection result was an assessment of three different predictors. During the feature selection, at each iteration, BFE used 60\% randomly selected samples for training and the rest 40\% samples to test the model. 

\subsubsection{Pattern discovery}
\label{subsec:pattern_des}
In general, the primary aim of self-organizing map (SOM) is to map $ m $-dimensional data onto a 2-dimensional (2D) plane. The 2D plane of SOM consists of a network of neurons (nodes). The network's nodes acquire the underlying property of the input data samples (e.g., events in the environmental data). Moreover, a SOM projects similar data samples to a cluster center (a node in a SOM) as per the similarity (Euclidean distance) of the data sample to the node~\cite{kohonen1990,vesanto1999}. 


SOM is an appropriate choice for this problem since it is tedious to define the number of clusters, especially when problems have complex relations between the features. SOM produced clusters automatically (see Section~\ref{subsec:pattern_des_RES}). {\color{black}Additionally, to analyze pattern related to the geo-locations}, geo-locations referenced mean physiological response $ r_{\text{mean}i} = (r_{x_i,y_i}^{p_1} + r_{x_i,y_i}^{p_2}+ \ldots +r_{x_i,y_i}^{p_N})/N $ across all participants was computed by matching GPS location information ($ x_i$: latitude, $y_i $: longitude) and aggregating the samples. Geo-location referenced mean physiological responses $ r_{\text{mean}i}  $ were computed to visually understand patterns in participants' physiological responses related to the actual map of the neighborhood, described in Section~\ref{subsec:pattern_des_RES}. 

\section{Results}
\label{sec:results}

\subsection{Sensitivity analysis (non-inferential modeling)}
\label{subsec:non_inf_RES}
\begin{figure}
	\centering
	\includegraphics[width=0.8\columnwidth]{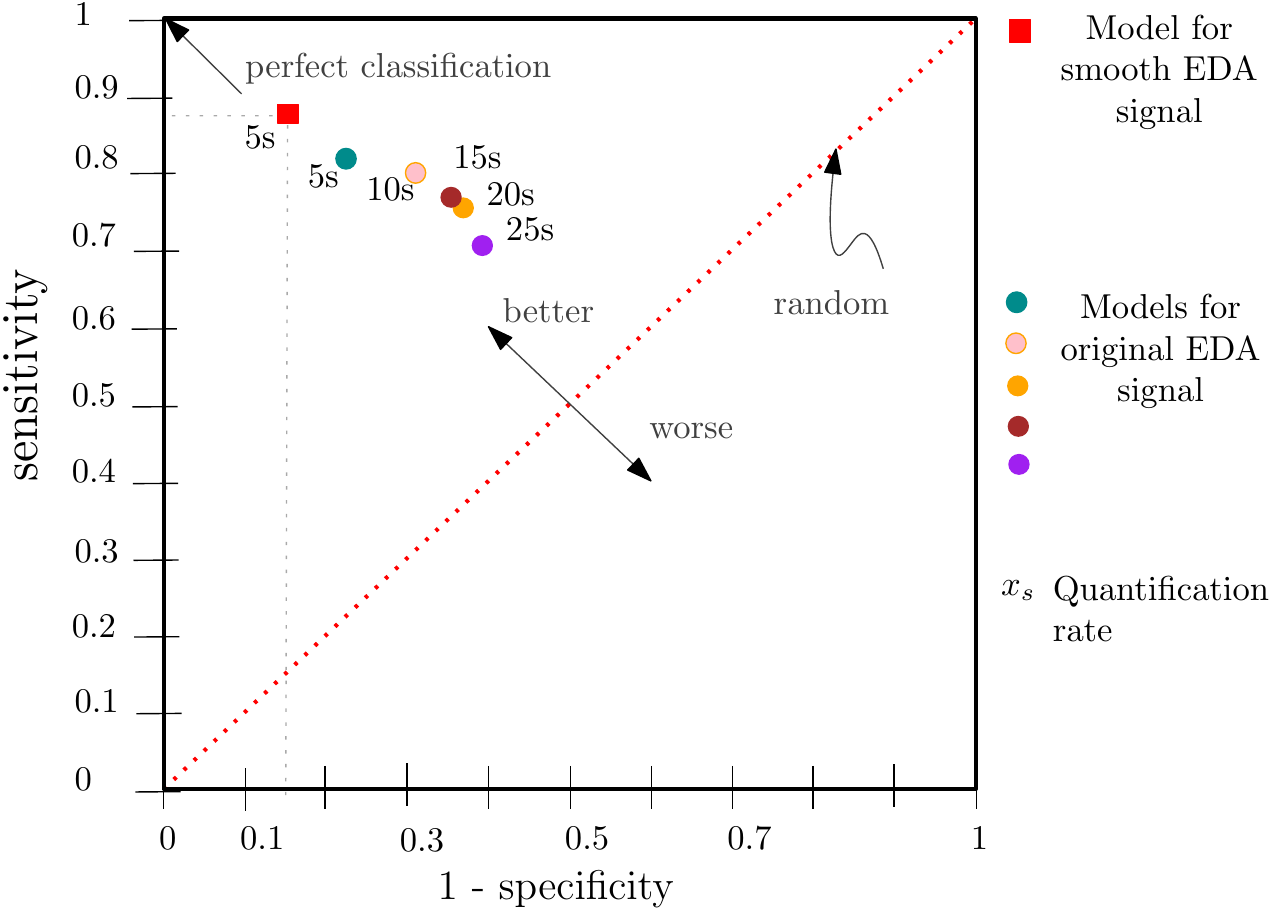}
	
	\caption{ROC graph of classification models on two categories of datasets represented in two different shapes: square and circles. Square represents dataset prepared with the output feature being the quantified smoothed EDA data; circles represent dataset prepared with the output feature being the quantified original EDA data.}
	\label{fig_roc}
\end{figure}

First, a classifier (REP-Tree described in Section~\ref{subsec:non_inf}) was trained and tested on the five ``time-resolved'' datasets corresponding to five quantification rates 25, 20, 15, 10, and 5 seconds, whose {\color{black}outputs} were labeled as the binary class: normal physiological response{\color{black},} ``N'' and aroused physiological response{\color{black},} ``A.'' The parameter settings used to train the REP-Tree models is in Table~\ref{tab_param}. The performances of the trained REP-Tree models are shown on a receiver operating characteristic (ROC) curve plot~\cite{pearce2000} in Fig.~\ref{fig_roc}. 

The model's performance improved as the quantification rates decreased (Fig.~\ref{fig_roc}). The model's high predictability for smaller quantification rates is an indicator of the participants' strong sensitivity towards the changes in the urban environment. The model's performance for smoothed EDA data (red square) was better than the model's performance for {\color{black}raw} EDA signal (circles). Thus, the smooth EDA data more accurately draw the association between a change in environmental features and participants' physiological states of arousal. 

The results of the 10-fold CV training of the RET-Tree classifier for both binary and multiclass classification for the dataset where smooth EDA data {\color{black}were} quantified at 5-second time-window as shown in Table~\ref{tab_rep_res}.  The classifier's predictive accuracy was found to be 87\% for the binary-class classification and 80\% for the multiclass classification.  

\begin{table}
	\centering
	\renewcommand{\arraystretch}{1.2}
	{\footnotesize 	
		\caption{ Average test results of 10-fold CV training of RET-Tree classifiers for both binary and multi-class classifications.}
		\label{tab_rep_res}
		\begin{tabular}{cllcccccccc}
			\toprule
			Classification & Class & TP & FP & TN & FN & Recall & Precision & Sensitivity & Specificity & Accuracy \\
			\midrule
			 \multirow{2}{2.5cm}{Binary class {\color{black}model}} 
			 & N & 3105 & 405 & 2162 & 396 & 0.89 & 0.88 & 0.89 & 0.84 & \multirow{2}{*}{87\%} \\
			 & A & 2162 & 396 & 3105 & 405 & 0.84 & 0.85 & 0.84 & 0.89 & \\
			 &  &  &  &  & &  &  &  &  & \\
			 \multirow{3}{2.5cm}{Multiclass {\color{black}model}}
			 & N & 3132 & 442 & 2125 & 369 & 0.89 & 0.88 & 0.89 & 0.83 & \multirow{3}{*}{80\%}\\
			 & LA & 1502 & 595 & 3392 & 579 & 0.72 & 0.72 & 0.72 & 0.85 & \\
			 & HA & 161 & 236 & 5346 & 325 & 0.33 & 0.41 & 0.33 & 0.96& \\
			\bottomrule	
			\multicolumn{11}{p{16cm}}{\textbf{Note:} For binary class, normal physiological response, ``N'' indicates nSCR = 0 and aroused physiological response, ``A'' indicates nSCR $ > $ 0. For multiclass, ``N'' indicates nSCR = 0; ``LA'' indicates a low arousal response, i.e., 0 $ < $ nSCR $ \le $ 6, ``HA'' indicates a high arousal response, i.e., nSCR $ > $ 6. The variables TP, FP, TN, and FN indicate true positive, false positive, true negative, and false negative, respectively~\cite{pearce2000}.}
		\end{tabular}}
\end{table}	

\subsection{Sensitivity range analysis (inferential modeling)}
\label{subsec:inf_RES}
The non-inferential model indicates that the participants' physiological responses are sensitive to the environmental changes. Therefore, we build an inferential model to understand how environmental features influence participants' physiological responses. A fuzzy rule-based inferential model was built using FURIA whose parameter settings are mentioned in Table~\ref{tab_param}. We adopted a binary-class classification of nSCR, where nSCRs were categorized into two classes: normal physiological response{\color{black},} ``N'' and aroused physiological response{\color{black},} ``A.'' The FURIA algorithm offered an average test accuracy of 70.23\% after a 10-fold CV training. Such accuracy is notably high for the complex problem of understanding {\color{black}the humans'} perception of their urban environmental conditions. 

\begin{figure}
	\centering
	\subfigure[]
	{
		\includegraphics[width=0.21\columnwidth]{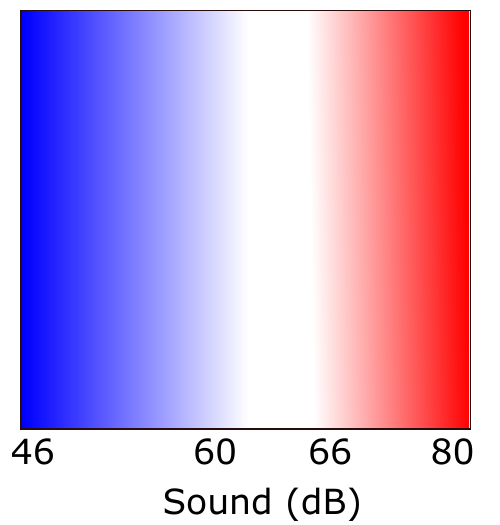}
		\label{fig_fr_sound}%
	}
	\subfigure[]
	{
		\includegraphics[width=0.23\columnwidth]{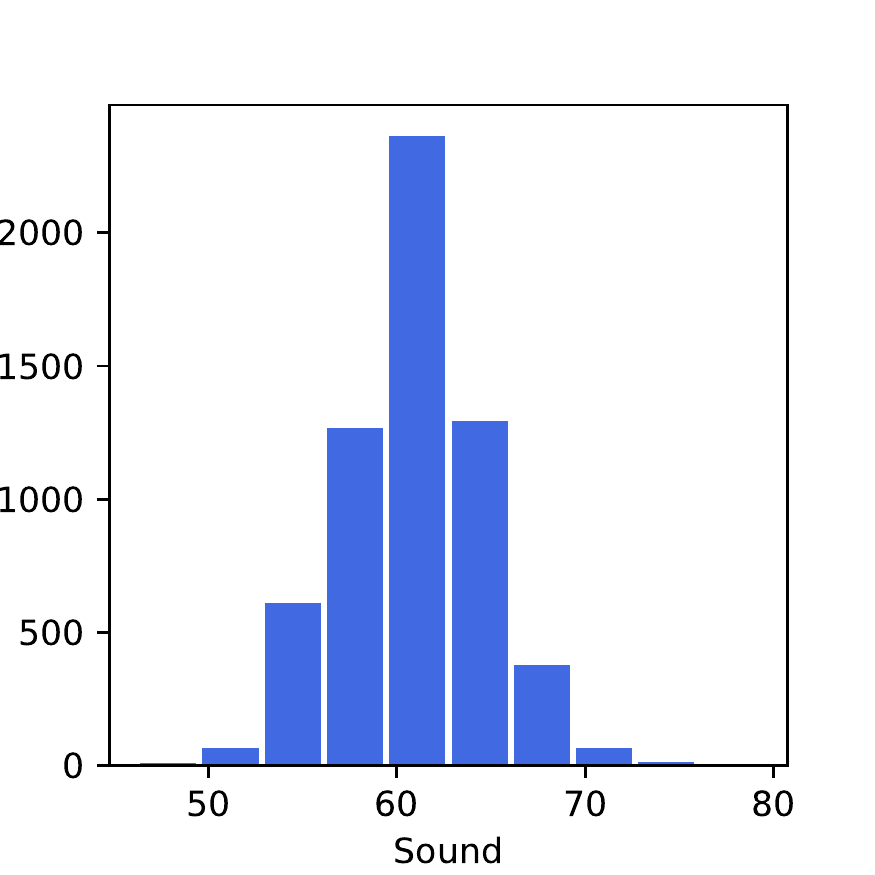}
		\label{fig_h_sound}%
	}
	\subfigure[]
	{
		\includegraphics[width=0.21\columnwidth]{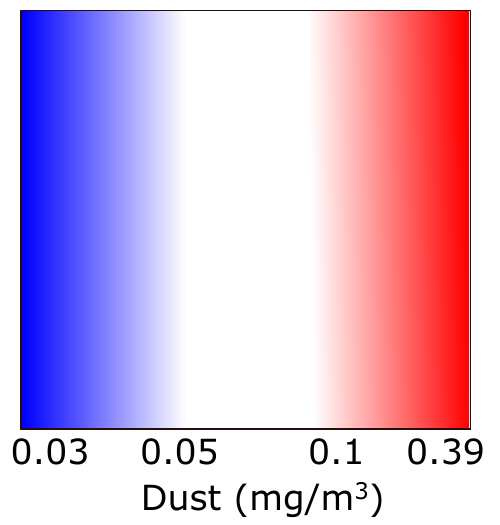}
		\label{fig_fr_dust}%
	}
	\subfigure[]
	{
		\includegraphics[width=0.23\columnwidth]{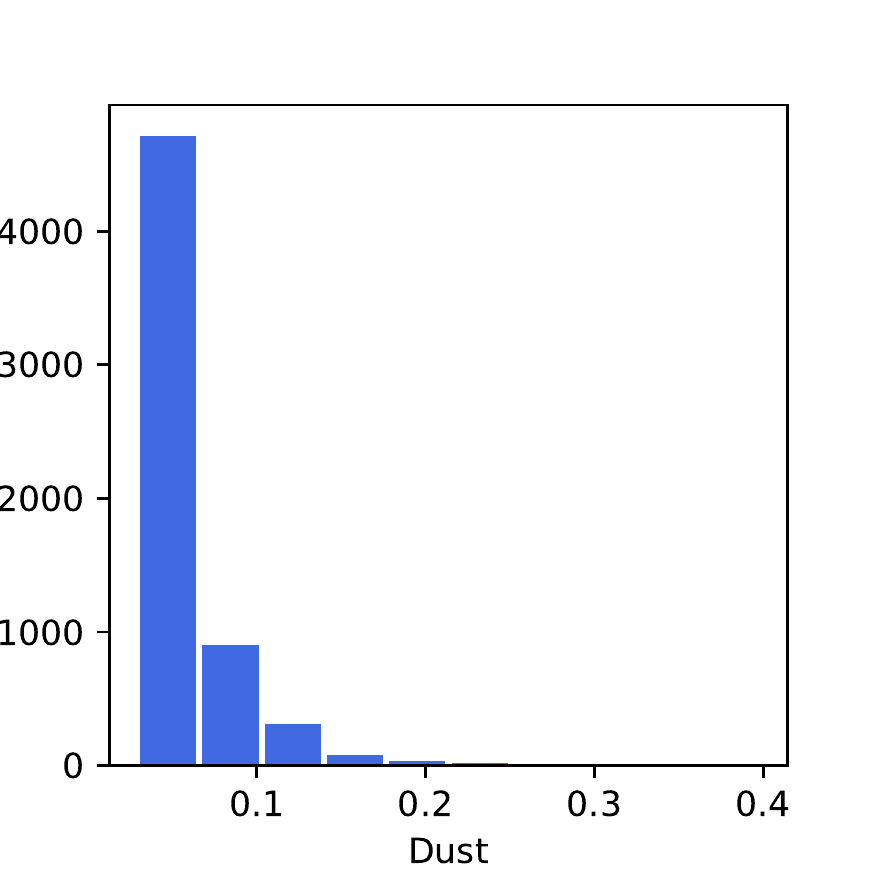}
		\label{fig_h_dust}%
	}
	\subfigure[]
	{
		\includegraphics[width=0.21\columnwidth]{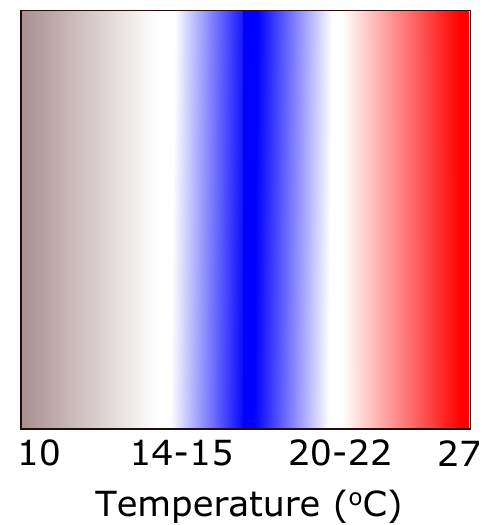}
		\label{fig_fr_temp}%
	}
	\subfigure[]
	{
		\includegraphics[width=0.229\columnwidth]{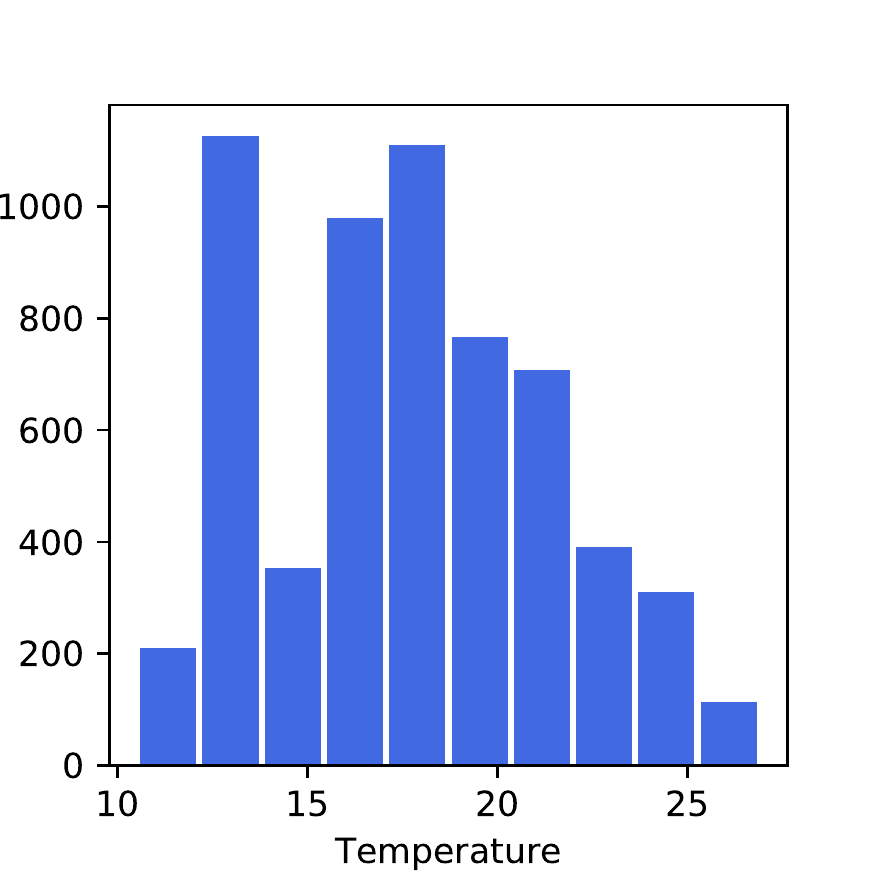}
		\label{fig_h_temp}%
	}
	\subfigure[]
	{
		\includegraphics[width=0.21\columnwidth]{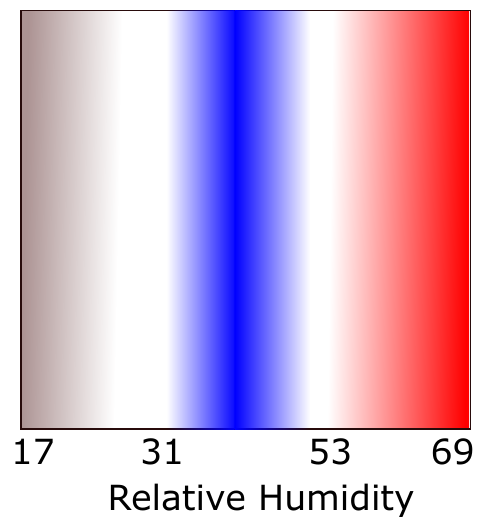}
		\label{fig_fr_rh}%
	}
	\subfigure[]
	{
		\includegraphics[width=0.23\columnwidth]{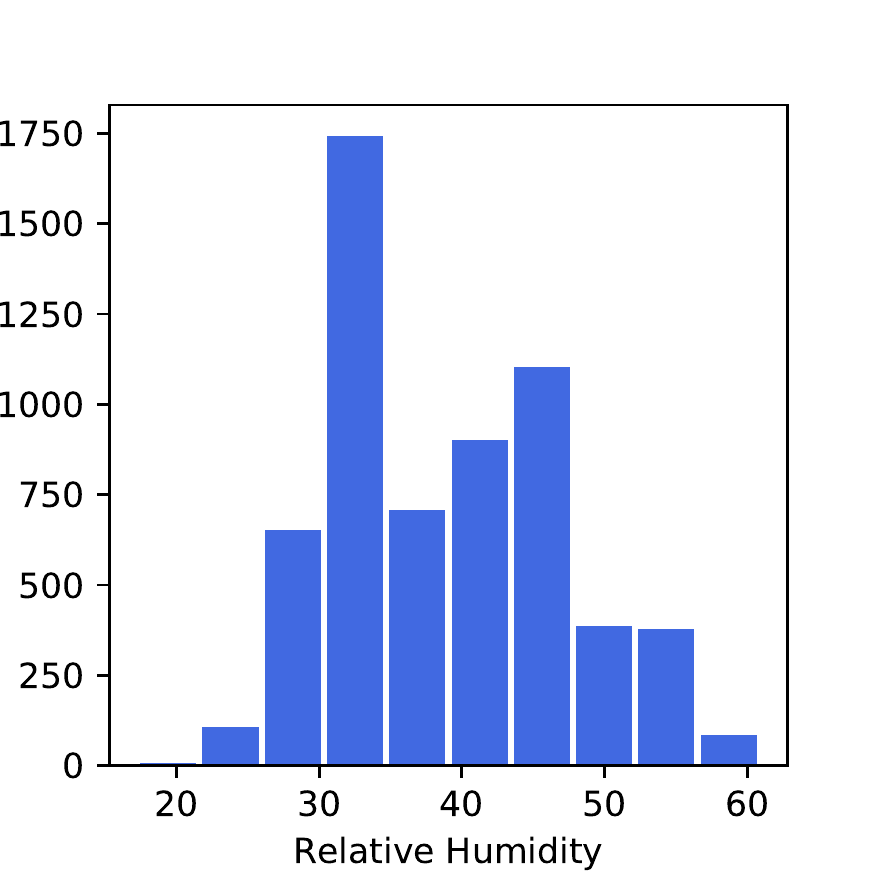}
		\label{fig_h_rh}%
	}
	\subfigure[]
	{
		\includegraphics[width=0.21\columnwidth]{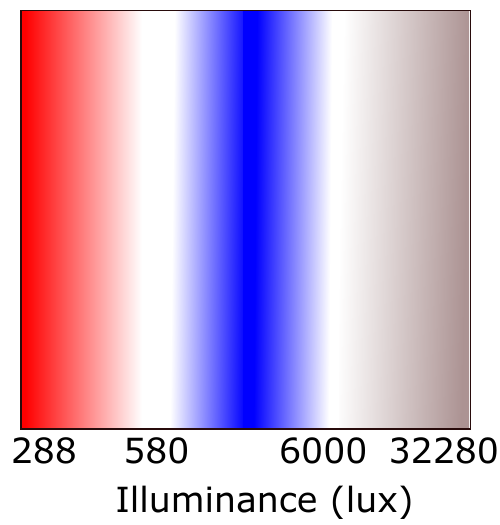}
		\label{fig_fr_light}%
	}
	\subfigure[]
	{
		\includegraphics[width=0.228\columnwidth]{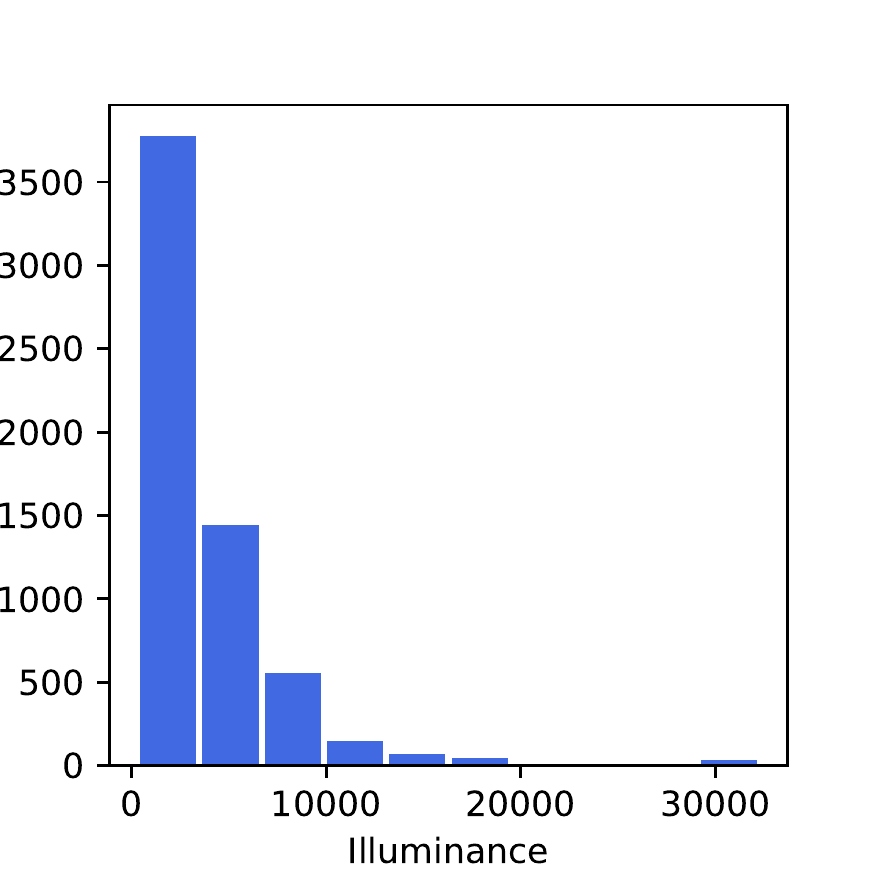}
		\label{fig_h_light}%
	}
	\subfigure[]
	{
		\includegraphics[width=0.21\columnwidth]{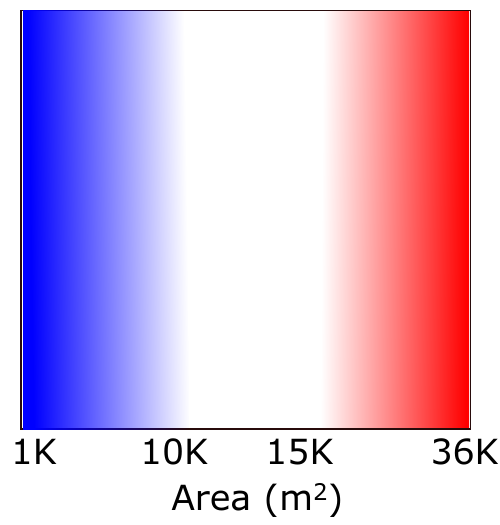}
		\label{fig_fr_area}%
	}
	\subfigure[]
	{
		\includegraphics[width=0.21\columnwidth]{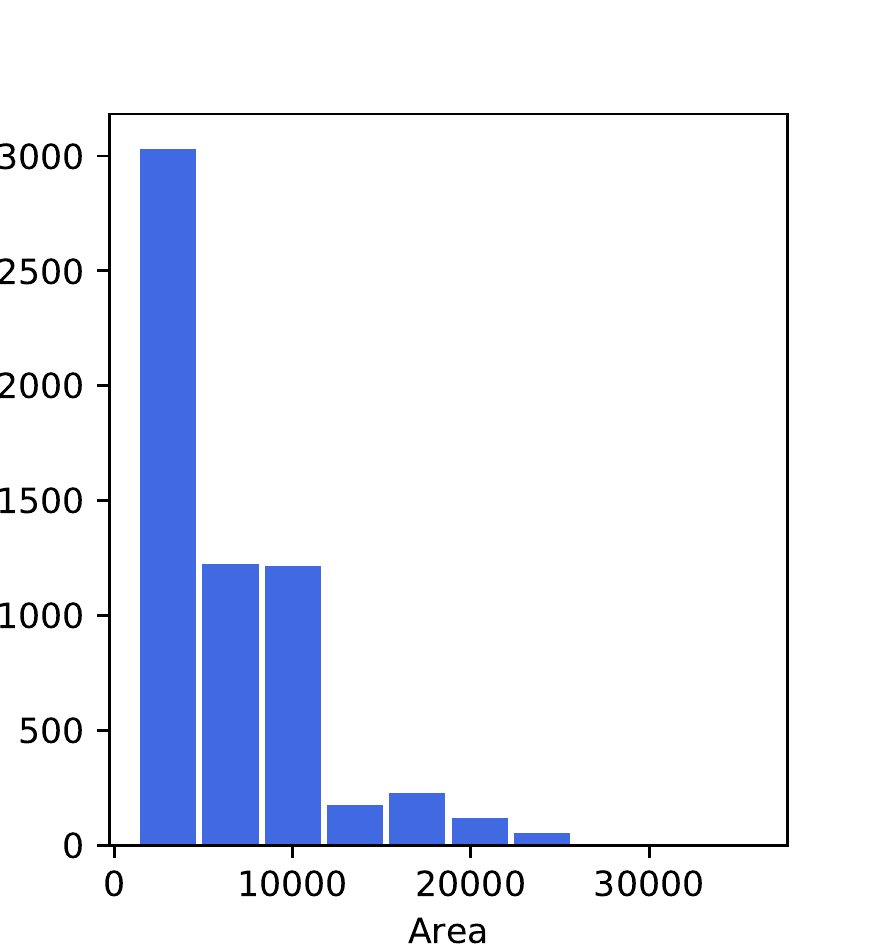}
		\label{fig_h_area}%
	}
	\caption{Visual interpretation of the fuzzy rules. The color ``red'' indicates the range for which the fuzzy rules finds nSCR $ > $ 0, i.e., an indicator of aroused physiological state. The color ``blue'' indicates the range for which the fuzzy rules finds nSCR $ = $0, i.e., an indicator of normal physiological state. The color ``white'' indicates a range of fuzziness. The color ``gray'' indicates the range for which rules do not provide any conclusive information.}
	\label{fig_fuzzy_rules}
\end{figure}

We analyzed the set of fuzzy rules generated by FURIA by segregating the rules between the participants' ``N'' and ``A.'' Fig.~\ref{fig_fuzzy_rules} is a visual interpretation of the obtained fuzzy rules for both classes ``N'' and ``A.'' We interpreted and represented the FURIA rules in Fig.~\ref{fig_fuzzy_rules} to find the values (range of values) of the environmental features that 
\begin{enumerate}[\quad(a)]
	\onehalfspacing
	\item were linked to class ``A,'' which indicates participants' aroused physiological state;
	\item did not significantly influence the participants' aroused physiological state.
\end{enumerate}


To validate the knowledge obtained from the visual interpretation of fuzzy rules, distributions of the environmental features were examined through histograms in Figs.~\ref{fig_h_sound}, \ref{fig_h_dust}, \ref{fig_h_temp}, \ref{fig_h_rh}, \ref{fig_h_light}, and~\ref{fig_h_area}. The visual interpretation {\color{black}and summarization} of the {\color{black}rules} for sound level in Fig.~\ref{fig_fr_sound} and its corresponding distribution in Fig.~\ref{fig_h_sound} indicate that the participants normal physiological responses match a particular sound level distribution. For example, the sound level distribution around \SI{60}{\dB} to \SI{66}{\dB} (Fig.~\ref{fig_h_sound}) correspond normal physiological state (Fig.~\ref{fig_fr_sound}). Furthermore, the participants had a tendency to exhibit aroused physiological state when experienced sound level above \SI{66}{\dB}. This result indicates that loud sound levels correspond to increased participant arousal.

The result was similar for temperature, where {\color{black}temperature degrees} greater than 21--22 $ ^\circ $C were associated with aroused physiological state (Fig.~\ref{fig_fr_temp}). However, it can be observed that the samples in the dataset for temperatures above 22 $ ^\circ $C were fewer than for the  {\color{black}temperature degrees} below 22 $ ^\circ $C (Fig.~\ref{fig_h_temp}), which we could take as confidence that heat alone did not cause the physiological arousal of participants. In (Fig.~\ref{fig_fr_light}), the participants exhibited physiological arousal for darker locations (illuminance {\color{black}level} below \SI{580}{\lux}).

\subsection{Simultaneous impact of environmental features}
\label{subsec:fs_RES}
\begin{figure}
	\centering
	\includegraphics[width=0.8\columnwidth]{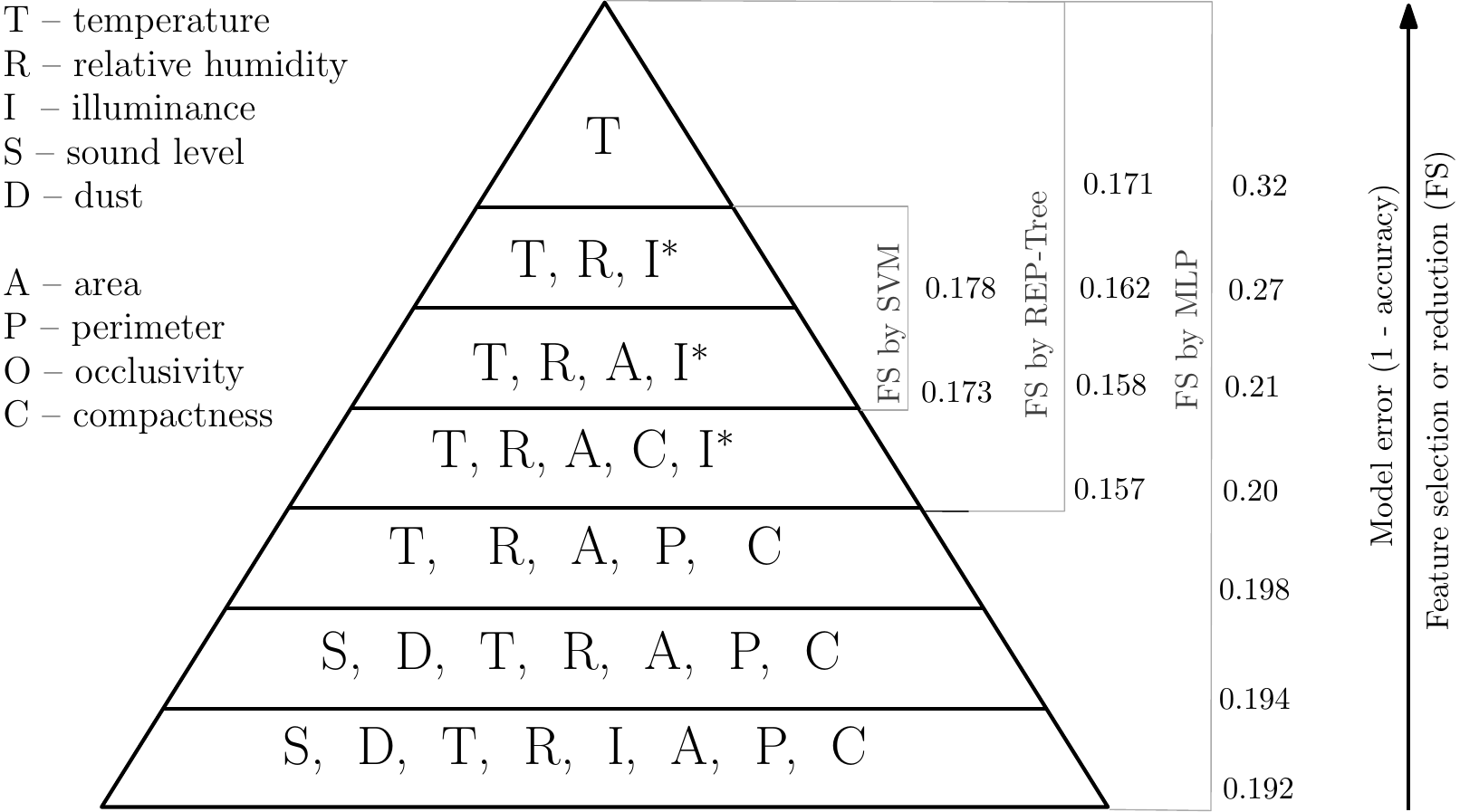}
	
	\caption{Hierarchy of feature importance. The symbol I* appeared only in the REP-Tree based feature selection. The feature set \{T,R,A,I\} appear in all three predictor's results.}
	\label{fig_feature_traingle}
\end{figure}

Inference modeling provided the values for environmental features {\color{black}that were} responsible for normal and aroused physiological states. However, it is also essential to discover which of the environmental feature(s) have {\color{black}the} strongest influence on the participants' physiological {\color{black}responses}. {\color{black}Hence}, we constructed a backward linear filter elimination (BFE) based feature selection framework and analyzed the obtained results to build a significance hierarchy of feature subsets (Fig.~\ref{fig_feature_traingle}). A feature subset's significance was estimated on its ability to predict ``N'' and ``A'' classes with high accuracy. 

Fig.~\ref{fig_feature_traingle} is a significance hierarchy triangle of the feature subsets, where a subset's predictability reduces when the number of features in the subset decreases. Three predictors provided three feature selection result sets. Fig.~\ref{fig_feature_traingle} is the compilation of the three result sets from all three predictors. The MLP, REP-Tree, and SVM agreed on the feature subset {temperature, humidity, illuminance, and Isovist area}, where the REP-Tree had the highest accuracy, followed by SVM and MLP. Therefore, temperature, humidity, illuminance, and Isovist area, {\color{black}were} noted as the most significant feature set but is a matter of trade-off between accuracy and number of features as indicated in hierarchy triangle (Fig.~\ref{fig_feature_traingle}).

\subsection{Patterns of perceptual variations}
\label{subsec:pattern_des_RES}
The predictive modeling confirmed the sensitivity of participants' physiological responses towards dynamic environmental conditions. The fuzzy rule-based analysis described the relationship between the environmental features and the physiological response. Feature selection indicated the most significant environmental features. However, pattern discovery {\color{black}explains}:

\begin{enumerate}[\quad(a)]
	\small
	\onehalfspacing
	\item which participants {\color{black}were} experiencing a similar {\color{black}environmental conditions and what were their response};
	\item {\color{black}whether} the participants' physiological responses for certain environmental conditions {\color{black}were similar};
	\item the patterns of the environmental features that influence the participants physiological arousal. 
\end{enumerate}

\begin{figure}
	\centering
	\subfigure[]
	{
		\includegraphics[width=0.4\columnwidth]{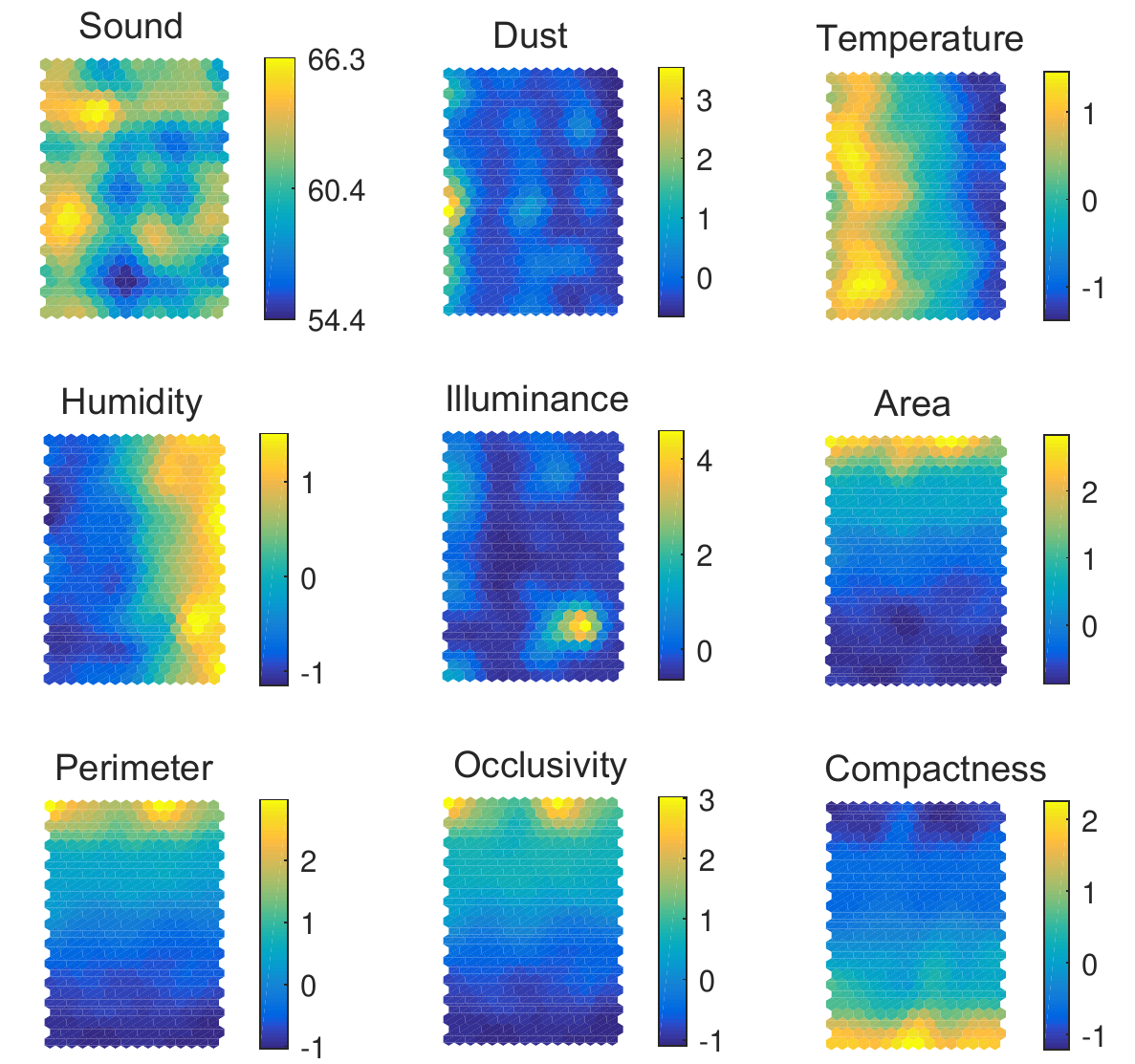}
		
		\label{fig_som_FM}
	}
	\subfigure[]
	{
		\includegraphics[width=0.27\columnwidth]{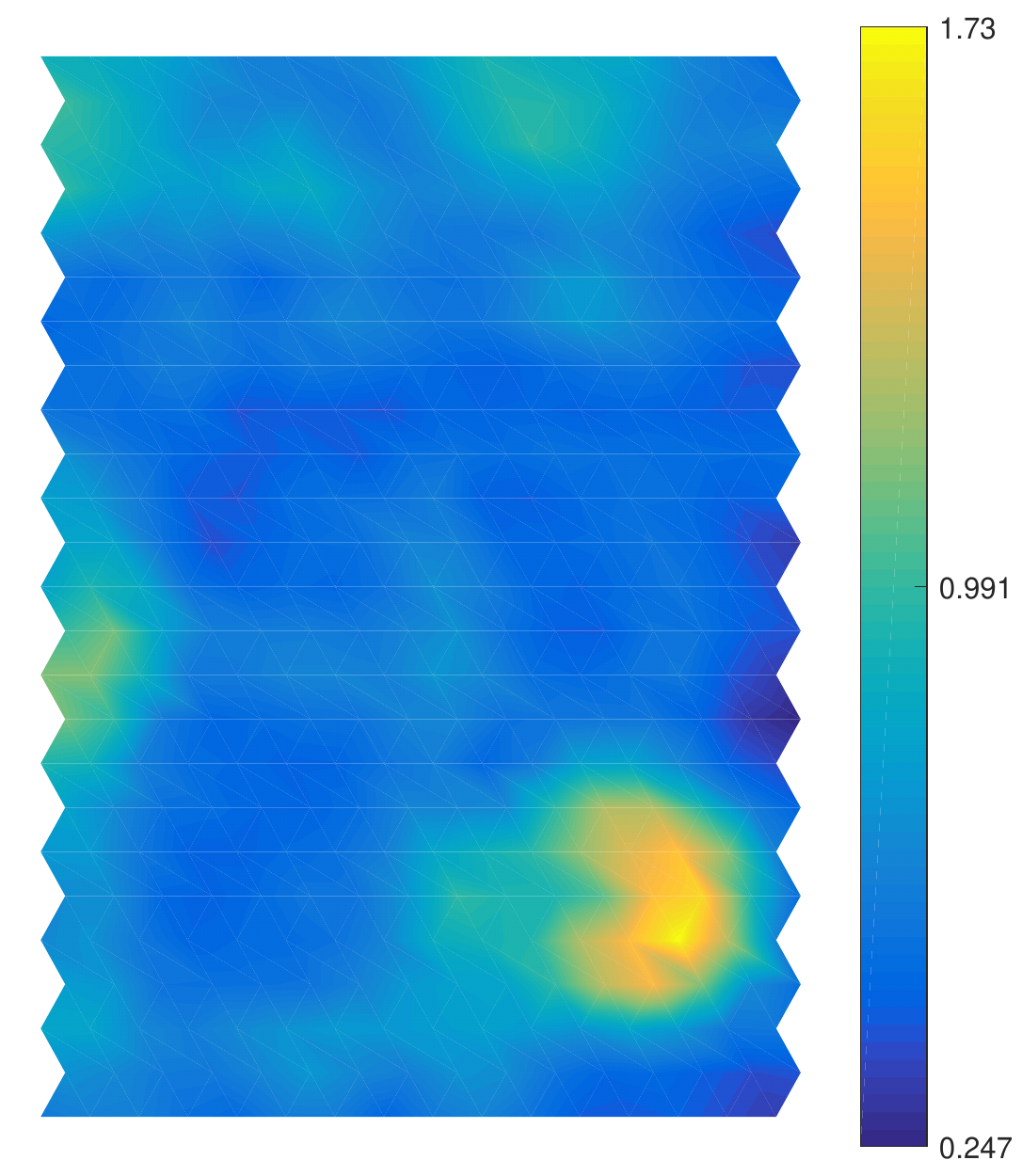}
		\label{fig_som_UM}%
		
	}
	\subfigure[]
	{
		\includegraphics[width=0.23\columnwidth]{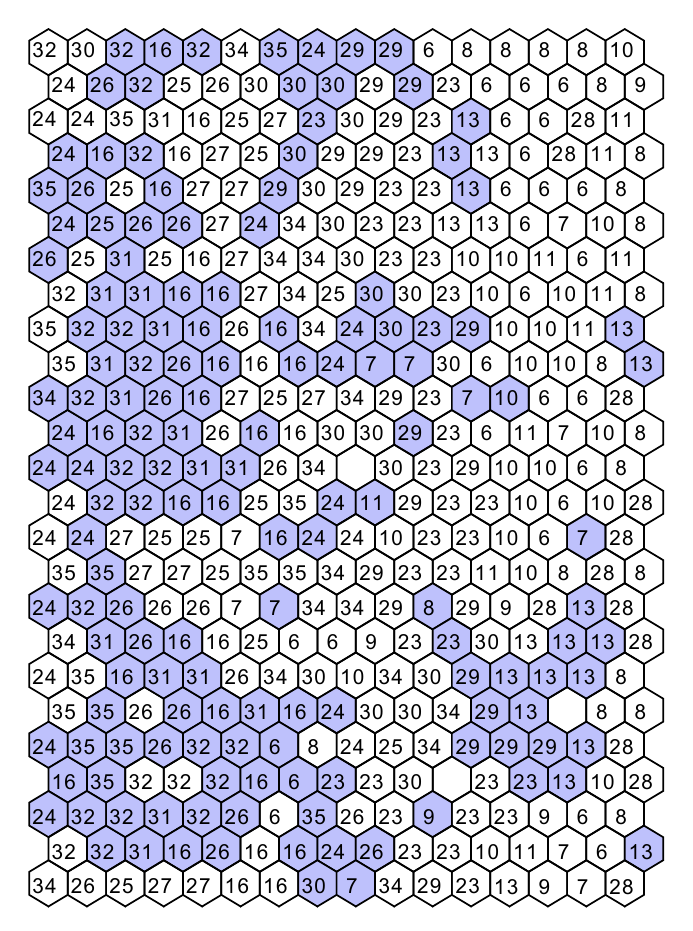}
		
		\label{fig_som_LM}
	}
	\caption{Trained SOM results; node value in the maps are indicated by color: lowest value is shown in dark blue, and the highest value is shown in bright yellow. (a) U-matrix: SOM clustering map. (b) F-matrix: maps for environmental features, which were linearly scaled with a variance of 1.0 so that they have equal importance in clustering. (c) L-matrix: participant ID and participants’ physiological response state label (``N'' and ``A'') map.}
	\label{fig_som}
\end{figure}

The compiled data (see Fig.~\ref{fig_main_framework}) were analyzed using SOM. Fig.~\ref{fig_som} is a result of automatic clustering from a trained SOM, where the 9-dimensional input data were mapped onto the $ 20\times20 $ dimension 2D plane consisting of hexagonal nodes. Each node in the map acquired the property of a set of samples.  Fig.~\ref{fig_som_FM} shows the maps of the environmental features on feature matrices (F-matrices). On a  feature matrix (F-matrix) of an environmental feature (e.g., sound level), the features' value assigned to F- matrix nodes are corresponding to the nodes on the SOM's unified distance matrix (U-matrix) in Fig.~\ref{fig_som_UM} and Label matrix (L-matrix) in Fig.~\ref{fig_som_LM}. Hence, the position and value of the nodes in all the maps (matrices) in Fig.~\ref{fig_som} are comparable to each other. More specifically, the U-matrix is the result of the F-matrices of the environmental features, and the L-matrix is the corresponding dominant label associated with the nodes. Therefore, to make sense of the pattern, we need to compare all matrices with one another.

The U-matrix in Fig.~\ref{fig_som_UM} shows the clusters of similar data points. The nodes with small differences (in terms of Euclidean distance) are shown in dark blue, and the nodes with high differences and are shown in bright yellow. In addition, the patches of nodes with similar colors, separated by lighter colors, indicate the clusters of data samples. Moreover, the data samples corresponding to a cluster in the U-matrix share a commonality, and dissimilar data samples are further apart. It is therefore implied that the {\color{black}participants'} ID label belonging to a cluster experienced similar environmental conditions. 

Fig.~\ref{fig_som_LM} is an L-matrix with each node was labeled with participant ID and the state of physiological response. White nodes indicate a normal physiological response and blue nodes indicate the aroused physiological response. By comparing these matrices, one can discover relevant patterns in the organization of the dataset. This could carefully be interpreted as a ``cause'' (Fig.~\ref{fig_som_FM}) and ``effect'' (consult with Fig.~\ref{fig_som_UM} and Fig.~\ref{fig_som_LM}) of the dynamic and simultaneous environmental features with the participants' physiological responses.

On the U-matrix (Fig.~\ref{fig_som_UM}) a bright yellow patch separates itself from all the other nodes clusters. This distinctly available yellow spot is the result of a high concentration of a set similar input samples, which in this case, is due to the concentration high illuminance values as evident from F-matrix for illuminance (Fig.~\ref{fig_som_FM}). Fig.~\ref{fig_som_LM} shows that at the exact same spot, participants' had aroused physiological state (most of the nodes are colored blue) and nodes were labeled with participants ID's (8, 13, 23, and 29) indicating that all the participants exposed to extremely high illuminance also experienced an equal aroused physiologically state. 

Additionally, three other clusters of dark blue exist on the U-Matrix in Fig.~\ref{fig_som_UM}: one at the bottom-left, one at the top-left and one at the top-right above the yellow patch. Investigating the F-matrices in Fig.~\ref{fig_som_FM}, we can find that the clusters at the bottom-left and the top-left in Fig.~\ref{fig_som_UM} are the results of high values of sound and temperature and extremely low values of illuminance. These clusters, when compared to L-matrix in Fig.~\ref{fig_som_LM}, indicate that the majority of participants responded with an aroused physiological state. 
Similarly, the cluster on the top-right is due to a combination of low values of dust and temperature. The corresponding L-matrix in Fig.~\ref{fig_som_LM} has the majority of nodes indicating a normal physiologically state. Further, the F-matrix for Isovist area in Fig.~\ref{fig_som_FM} shows that the high value of Isovist area resulted in an aroused physiological state, also evident from the L-matrix in Fig.~\ref{fig_som_LM}. L-matrix also indicates that participant IDs 16, 23, 24, 29, 32, and 35 experienced such a high Isovist area and responded with a  similar physiological state. 

\begin{figure}
	\centering
	\includegraphics[width=\columnwidth]{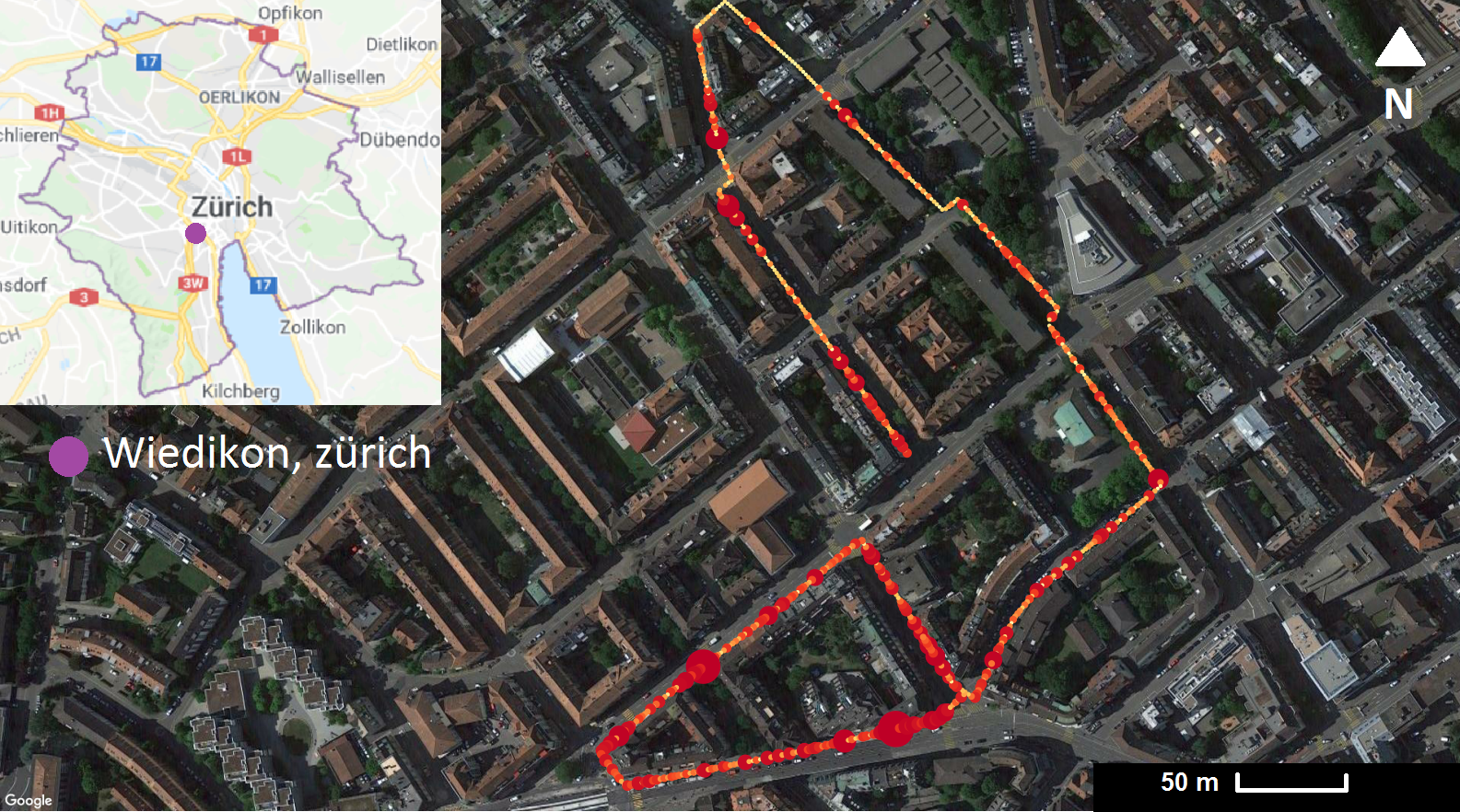}

	\caption{Geo-location referenced mean physiological responses across all participants. An animation of {\color{black}this graphic} indicating real-time simulation is available at~\cite{esum2018}.}
	\label{fig_geo}
\end{figure}

In pattern analysis, the mean physiological response across all participants was mapped onto the geographic location along the path. The geo-location referenced mean physiological response was computed and normalized between 0 and 1. The geo-location referenced physiological responses highlighted specific locations on the neighborhood's map where participants experienced aroused physiological state (Fig.~\ref{fig_geo}). The locations, where on average all participants exhibited high physiological arousal response are indicated in red while low physiological arousal is indicated by yellow. Varying size of dots on the map in Fig.~\ref{fig_geo} is proportional to the degree of participants' physiological arousal.

\section{Discussion}
\label{sec:disc}
Through this research, we extracted patterns from the data gathered during a controlled {\color{black}study}, where we asked participants to walk through an urban environment (Section~\ref{subsec:study}). Our data analysis methods had the following dimensions: signal processing, multi-sensor information fusion, and knowledge mining using machine learning techniques. The sensor frequency unification and quantification led to the preparation of identically independent data samples of events and corresponding physiological response. During the data processing phase, we categorized physiological response data (EDA signals) into clean and erroneous signals (Section~\ref{subsec:data_clean}).
EDA signal recording is susceptible to artifacts and the suggested definition identifies an erroneous EDA signal. Finally, the quantification method segmented the continuous temporal data into regular time intervals of {\color{black}$ t $-seconds ( time-window size) and the} quantification rate of 5-seconds was most efficient (Section~\ref{subsec:non_inf_RES}). 

We applied both supervised and unsupervised machine learning techniques. This included testing REP-Trees' predictive accuracy {\color{black}in determining the models' sensitivity towards five different quantification rates: 5, 10, 15, 20, and 25 seconds.} The predictive model at $ t=5 $ seconds had the highest accuracy (Fig.~\ref{fig_roc}). 
The high accuracy of the REP-Tree model indicates its predictive ability of the participants' normal and aroused physiological responses state for a given set of environmental condition consisting of sound level, dust, temperature, humidity, illuminance, and Isovist descriptors. 

The inference modeling, in addition, produced exact values of the environmental features and their influence on participants physiological response state (Section~\ref{subsec:inf_RES}). Also, the environment features with the largest range of values in the dataset (highest distribution) were directly linked to normal physiological responses (Fig.~\ref{fig_fuzzy_rules}). In other words, the participants showed a ``habitual effect,'' and they tend to respond differently to a change in the environmental condition from the previous one (Section~\ref{subsec:inf_RES}). 
Such a generalization of fuzzy rules across all participants is limited because of {\color{black}the} availability of {\color{black}fewer} data samples and the variations in cities' architectures. However, it is necessary to mention that all participants engaged in the study on different days and different time-of-day and we observed a high accuracy in the model's predictability despite being applied to such a complex and diverse dataset. For example, a fuzzy rule (Fig.~\ref{fig_fuzzy_rules}) indicates that a participants' arousal levels correspond to extremely low illuminance, or high temperature, or a large Isovist area. Specifically, change in physiological arousal was observed for a small to a large Isovist area, {\color{black}i.e.,} an entry to a crossroad and passing from a narrow to a wider street (Fig.~\ref{fig_geo}).    

It was difficult to identify the features with the highest influence on {\color{black}the} physiological response from the inference modeling. Therefore, a backward feature elimination method with three predictors (Section~\ref{subsec:fs}) helped {\color{black}determine} the most significant environmental feature(s) and is presented in a significance hierarchy triangle (Fig.~\ref{fig_feature_traingle}). The feature selection process, however, had its trade-off; when reducing the number of features from the feature set, it also decreases the accuracy of the predictors. After a thorough inspection of Fig.~\ref{fig_feature_traingle}, the predictors suggest the temperature, humidity, illuminance and the Isovist area as the most significant features set  compared to a set of any other features combination (Section~\ref{subsec:fs_RES}). 

SOM was employed for automatic clustering to discover patterns in the dataset  (Section~\ref{subsec:pattern_des_RES}). The participants with similar environmental conditions were expected to have a similar perception (physiological arousal state) and expected to fall into {\color{black}the} same cluster or node on the map. For example, a cluster formed due to extremely high illuminance and another for low illuminance conditions (Fig.~\ref{fig_som}). This indicates that a particular environmental condition influences most of the participants equally and the majority of participants responded a similar physiological response state when experiencing similar conditions. 
Furthermore, because the participants walked at different speeds, the number of quantified events corresponding to each participant slightly varied. Therefore, the geo-location referenced normalized mean of the events was the best method to show the geolocation of the participants' average physiological responses on the map (Fig.~\ref{fig_geo}). This map can be used to visually inspect the impact of urban features, such as street-width, street-type, traffic, type of area (residential and industrial) and their potential impact on the participants' physiological response. 

\section{Challenges and opportunities}
\label{sec:challanges}
The methods developed for this investigation help reveal patterns from complex human-environment interactions.  The analysis predominantly focused on improved quantification methods for physiological arousal level detection and a means to correlate arousal level with environmental stimuli.  This approach allows us to observe an increase in physiological arousal in response to specific environmental conditions (Section 5).  The primary challenge of this study was the process of selecting the appropriate tuning parameters to quantify and evaluate the arousal label.  For example, the accuracy of the methods (Fig.~\ref{fig_roc}) varied depending upon the quantification rate.  Similarly, the accuracy of the method depends on the procedure and threshold adopted for the nSCRs level detection~\cite{braithwaite2013}. Moreover, we captured 9 features of a real-world dynamics situation. Hence, increased number of features may further improve the predictive model's accuracy. 

Future studies can utilize the presented experimental design and quantification methodology. For instance, it can be extended to capture citizen's public transport commuting experience (physiological response while walking, waiting, and riding), and for traffic safety, the method can be potentially applied to understand the physiological arousal pattern of vehicle riders while they ride through cities~\cite{collet2003,shuyun2009}. Moreover, the developed predictive model can be used to extrapolate the potential citizen's arousal levels to a larger geographic area when combined with the isovist values and measured environmental data beyond the selected path. 

{\color{black}
In this research, we recognized factors influencing humans perception. Whereas to meet the refereed challenges, our findings suggest that further employing virtual reality set-up could help reducing noise that may be induced by unknown factors. Additionally, our findings suggest that a subjective thresholding skin conductance can also be employed to mitigate the challenges.

Moreover, in the field of urban studies, it is crucial to understand how the built environment influences human behavior and perception. This question has been central to the practice and research ever since and poses a fundamental methodological problem since it is especially difficult to a) objectively measure perception and b) deal with the multitude dynamic environmental factors preventing to identify the effect of pure urban form on human perception. As an answer to this problem, this research provides a major contribution by presenting and empirically testing a novel research framework for predicting and inferring the effects of planning decisions on human perception. In essence, the framework provides insides into How, and Why do architecture and urban design influence human perception which is particularly helpful for evaluating planning proposals and to guide the design decisions. For this purpose, we adopt the state of the art mobile sensing technologies as well as machine learning methods which are specifically chosen and adapted for needs of architecture and urban design research.}

\section{Conclusions}
\label{sec:con}
This research presented a specific methodology to evaluate a complex dataset  from an experiment with physiological responses of 30 participants linked to environmental conditions. The measurements in the dataset came from seven sensors with differing frequencies and four additional geometric features. The proposed {\color{black}data} quantification and multi-sensor information fusion methods linked participants' physiological state of arousal to environmental conditions. Four categories of machine learning techniques (non-inferential modeling, inferential modeling, feature selection, and clustering) {\color{black}revealed} patterns in the dataset: The high accuracy of the non-inferential predictive model was an evidence of the participants' physiological state sensitive to the changes in environmental conditions. The fuzzy rule-based inferential modeling results indicate that the occurrence of ``normal'' and ``aroused'' physiological conditions corresponds to specific values (and range of values) for each environment feature. It suggested that the changes in the participant physiological arousal state primarily occurred due to the fluctuations in the environmental conditions. Feature selection showed that some environmental features, such as temperature, humidity, illuminance, and the-filed-of-view were more dominant in their influence on participants' physiological response than sound level and dust. Pattern analysis from self-organizing map indicated that, primarily, the participants who experience similar environmental conditions responded in similar physiological arousal state. Finally, the geo-location referencing of average physiological response across all participants produced a means to visually inspect how participants respond during the actual walk in relation to permanent urban features. The proposed data analysis framework {\color{black}revealed} patterns from the complex spatial-temporal environmental and physiological data that impact our understanding of urban settings.  

\section*{Acknowledgments}
This research was funded by Swiss National Science Foundation (SNF) project no. 100013L 149552 titled  under the German Research Foundation (DFG) Research Grant no. DO551/21-01​. Authors are thankful to all the participants who took part in the study. 

\end{doublespace}

\begin{onehalfspace}

\end{onehalfspace}

\appendix
\section{Appendix}
\label{sec:appen}
%
%


\setcounter{table}{0}
\renewcommand\thetable{\Alph{section}.\arabic{table}}
\begin{table}[!h]
	\centering
	\renewcommand{\arraystretch}{1.2}
	{\footnotesize 	
		\caption{Parameter settings of the machine learning techniques.}
		\label{tab_param}
		\begin{tabular}{llll}
			\toprule
			Algorithm/Tool & Parameter & Definition/Purpose & Value \\
			\midrule
			\multirow{5}{*}{Ledalab} & Analysis type & Type of method for decomposing a signal & Continuous decomposition  \\
			& Optimization time & Number of times a signal is optimized & 2 \\
			& Window range &  & 1 –3.7 Sec.\\
			& Smooth method  &  & Gaussian \\
			& Smoothing wind &  & 0.2 Sec. \\
			&  &  &  \\
			\multirow{3}{*}{REP-Tree} & \#Leaf instances & Minimum children per node. & 2 \\
			& Depth & Maximum limit of tree depth/level. & No limit \\
			& Pruning & Pruning of tree nodes. & True \\
			&  &  &  \\
			\multirow{1}{*}{FURIA} & Function & Membership function for fuzzification & Trapezoidal \\
			&  &  &  \\
			\multirow{4}{*}{MLP} & Learning rate & Convergence speed. & 0.3 \\
			& Momentum rate & Influence of previous iteration. & 0.2 \\
			& Hidden Layer & Maximum hidden layer nodes. & 10 \\
			& Iterations & Maximum time for parameter optimization  & 1000 \\
			&  &  &  \\
			\multirow{2}{*}{LibSVM}	& SVM kernel & Type of function at a hidden node & Radial basis function \\
			& Epsilon & Termination criteria for algorithm & 0.001 \\
			&  &  &  \\
			\multirow{5}{*}{SOM}& Map dimension & Dimension of the 2D plane &	20x20 \\
			& Normalization & Method of data normalized for SOM training & Linear scaling \\
			& Training mode & Number of samples in an epoch of training. & Batch \\
			& Iteration & Number of training epochs & 25\\
			& Fine-tuning & Number of fine-tuning epochs & 20\\
			\bottomrule
		\end{tabular}}
\end{table}			
\end{document}